\documentclass[a4paper,twoside,11pt]{article}
\usepackage[francais]{babel}
\usepackage{graphics,graphicx}
\usepackage{latexsym,amsfonts}
\usepackage{amssymb}
\date{}
\title{\textbf{An efficient classification in  IBE}\\
\textbf{Provide with an improvement of BB2 to an efficient
Commutative Blinding scheme } }
\author{Rkia Aouinatou$^{1}$,  Mostafa $Belkasmi^{2}$ \\
 $^{1}$
\normalsize Faculty of Sciences, Mohamed V-Agdal B.P. 1014 Rabat, Morocco \\
$*$ \textit{\small Laboratoire de Recherche Informatique et Telecommunication: LRIT }\\
\textit{\small \textbf{Email:}} \small rkiaaouinatou@yahoo.fr\\
\normalsize $^{2}$ ENSIAS: University
 Mohammed V- Souissi, Rabat, Morocco \\
\textit{\small \textbf{Email:}} \small belkasmi@ensias.ma}
\begin{document}
\maketitle
\begin{center}
\section*{Abstract}
\begin{flushleft}
 Because of the revolution and the success of the technique IBE
(Identification Based Encryption) in the recent years. The need is
growing  to have a standardization to this technology to
streamline communication based on it. But this requires a thorough
study  to extract the strength and weakness of the most recognized
cryptosystems. Our first goal in this work is to approach to this
standardization, by applying a study which permit to extract the
best cryptosystems.\\
As we will see in this work and as Boneh and Boyen said in 2011
(Journal of Cryptology) the BB1 and BB2 are the most efficient
schemes in the model selective ID and without random oracle (they
are the only schemes traced in this model). This is right as those
schemes are secure (under this model), efficient and useful for
some applications. Our second goal behind this work is to make an
approvement in BB2 to admit a more efficient schemes. We will
study the security of our schemes, which is basing on an efficient
strong Diffie-Hellman problem compared to BB1 and BB2. More than
that our HIBE support $s^{+}$ID-HIBE compared to BBG (Boneh Boyen
Goh). Additionally the ID in our scheme will be in $Z_{p}$ instead
of ${Z_{p}}^{*}$ as with BBG. We will cite more clearly all these
statements in in this article.
\end{flushleft}
\end{center}
\section*{ keywords} \normalsize  IBE, competition, RO, SM, sID,
BF, SK, BB1, BB2, Water, Gentry, Problem Bilinear of Diffie
Hellman,  HIBE, BBG, selective ID, $selective^{+}$ ID,
$Z_{p}^{*}$, complexity, security.

\section{INTRODUCTION}
 IBE was proposed by Adi Shamir in 1984 [1] as a solution to the
problem of the revocation of the public key and the requirement of
the certificate in PKI. In IBE (Identification-Based Encryption)
the public key can be represented as an arbitrary string such as
an email address. It's corresponding private key is generated by a
Private Key Generator (PKG) who authenticate users according to
their corresponding identities. This idea was proposed by Shamir
only as concept. And we will wait until 2001 at which Dan Boneh
and Mathew Fanklin [2] propose an elegant scheme in the Random
Oracle, using the pairing. Their proposition open the door to a
more efficient scheme (with pairing), we cite: Boneh-Franklin (BF)
[2], Skai-Kasarah (SK) [3] under the model Randoms Oracles,
Boneh-Boyen (BB) [4] under the model selective ID, Water [5] and
Gentry [6] under   Standard Model. These cryptosystems are the
great themes of the cryptography IBE, because
all the cryptosystems  which comming later: [7,8] and others, are just their modified.\\
After all these proposals several companies have begun working
with  IBE instead of the PKI. We can cite Voltage Security and
Nortech. This seeks to balance the standardization of the
communication, which is currently being prepared (already tried by
IEEE [9]). But to do it we need a very thorough study because we
need to consider many things. In this study we make a comparison
between the main cryptosystems we have cited.\\ The comparison in
the IBE has been treated in a lot of papers, for example:  Boyen
[10]
 call to the standardization of BB1 (IEEE 1363.3) by showing
its benefit. The same author [11] make a comparison between BB1,
SK, BF. In [7] Kiltz-Vahl propose  two cryptosystems which they
have shown their advantage over that of Gentry and Kiltz-Galindo.
Note that every time a cryptosystem is invented it begins to
describe their advantages over others. Unfortunately all these
studies are not conclusive. Because, either  they do not take into
account all the major cryptosytems, or  the numbers of the factors
at which the comparison is based are insufficient. In this work we
will make a practical comparison between all the proposed
cryptosystems, by integrating the most possible factors and
proposing a suitable
 schedule.\\
Usually the systems networks become more accessible and open,
apparently  an active adversary (even passive) may not be limited
to eavesdropping, but may take a more active role. She can
interact with honest parties, she may analyze some older
responses,  she can try to break some problem of Diffie Hellman
used in the target cryptosystem...That's why it is out of habit
and within the cadre of standardization, that the security of each
cryptosystem will be checked by what is called studies of
simulations. Those studies  are introduced by [12], they are being
done in advance to test the rigidity of a cryptosystem. But all of
them require that the identity wishing to be attacked will be
asked in the challenge phase. We call this, full domain. In 2003
Canetti et al. [13] proposed a  weaker security model, called
selective identity IBE (sID-IBE). In this model the adversary must
commit ahead of time to the identity it intends to attack. In [14]
Sanjit Chatterjee et al have presented an extension of this model
at which the adversary is allowed to vary the length of the
challenge identity. Which is not allowed in the sID model.
Naturally any protocol secure in the $s^{+}ID$ model is also
secure in the s-ID model, but the
converse is not necessarily true.\\
Even if the reduction from selective-ID IBE to fully secure IBE
introduces a factor of N[4] (N will be at least $2^{160}$ to make
the problem bilinear rigid) in the security parameters of the
system. Boneh and Boyen in 2004 [4] have proposed tow efficient
schemes BB1 and BB2 under this model. The first one is in the
approach of Commutative Blinding, it is an HIBE scheme based on
the DBDHP (Decisional of Bilinear Diffie and Hellman Problem).
Until the second is in the Exponent-Inversion approach, it is an
IBE based on Dq-BDHIP (Decisional q-Invertible of Bilinear Diffie
and Hellman Problem).
\\
As an IBE requires the use of a PKG to generate the private key,
so  alone PKG is insufficient. Since, it will be a concentration
in one. To avoid this the works [15][16] and others are proposed.
All them are heavy, because  for k authority in hierarchy it
necessitate to generate k element in extract and in encrypt, in
addition to k product of pairing in decrypt. This cost was reduced
by Boneh, Boyen, Goh [17]. In [17]  the authors propose a scheme
where the ciphertext size and the decryption cost are independent
of the hierarchy depth. The ciphertexts is always just three group
elements and decryption requires two bilinear map computations.
This reduction influence on some application such as Forward HIBE
and Broadcast Encryption.\\ But even the authors in [17] reduce
the cost in the syntax of HIBE, their scheme requires that the
identity to be challenged will be in ${Z_{p}}^{*}$, because they
necessitate it in the technique of the study of simulation to
remove the master key $g^{\alpha}$. This limit the choice of the
identities  which is a restrict. More than that, their proposal
was not familiarized with the notion of $s^{+}ID$, and it is
proven in [14] that if want to convert s-ID to $s^{+}ID$ we will
make a degradation of h (h=v-$v^{+}$, v is the length of the
identity challenged, and $v^{+}$ is the target prefix).
This give a more advantage to attack the cryptosystem, as we may have an advantage equal to $h\varepsilon$.\\
Our second contribution behind this work is: To over come all
this. Keeping the syntax of BB2 (noting that BB1 and BB2 are
considered until 2011 [18] as an efficient schemes in the sID
Model),
 we will propose a scheme (with a little change in BB2) in the
 Commutative Blinding approach and
which requires only 1 pairing in decrypt contrary to BB1. With the
same manner, we will reduce the HIBE following BBG. This
reduction, will help us to give a more efficient Forward HIBE and
even Broadcast Encryption. By contrast to BBG our result HIBE
support $s^{+}ID$ model and it can project
 in the $Z_{p}$ contrary to  ${Z_{p}}^{*}$ as with BBG.
\section*{Organization}
Firstly we will divide our work in tow categories: First goal and
second goal.\\
We begging in the first goal by some preliminaries,  section
number 2.2 will be reserved to the comparison (in two level:
complexity and security). Our final decision will be given in
section 2.3.\\
For the second goal we will also staring by some notions, it
concerns the functionality of IBE, HIBE and their security, in
addition to that we give some preliminaries concerning the problem
of Diffie Hellman to be used. We reserve  section 3.2 and 3.3 to
our proposal for IBE and HIBE respectively, then we test the
efficiency of our schemes compared to BB1, BB2 and BBG. In section
3.4 we demonstrate the utility of our scheme for Forward scheme.
In the end we give a conclusion.
\section{First goal}
\subsection{Some Preliminaries}
Before giving some of these preliminaries, we remember that our
first goal about this work is to classify the main cryptosystems.
The cryptosystem's which are in competition are: Boneh Franklin,
Skai Kasarah, Boneh Boyen (BB1, BB2), Water, Gentry.
 \subsubsection{Relation of the Problems of Diffie and Hellman}
\subsubsection*{2.1.1-1 Problem Bilinear of Diffie Hellman}
 \textbf{Definition
1: (Bilinear Diffie Hellman Inversion Problem (k-BDHIP) [5])}. Let
 k be an integer, and x $\in$ $Z_{q}^{*}$, $ P_{2} \in G_{2}^{*}$, $P_{1}$ =
$\psi(P_{2})$, \^e : $G_{1} \times G_{2}$ $\longrightarrow$
$G_{T}$. Given $(P_{1}, P_{2}, xP_{2}, x^{2}P_{2},...,
x^{k}P_{2})$, compute
\^e($P_{1},P_{2})^{\frac{1}{x}}$ is difficult.\\
\textbf{Definition 2:New Problem: SiE-BDHP}
 (Simple Exponent
Bilinear Diffie Hellman Problem). We express it for the first time
in the literature: Let
 k be an integer, ($P_{1}$, $P_{2}$) in
$G_{1}$ $\times$ $G_{1}$, x$\in$$Z_{q}$, given $P_{0}$, $xP_{1}$,
x$P_{2}$, $xP_{3}$, $xP_{4}$, ..., $xP_{k}$. Compute
$xP_{0}$ is difficult\\
\textbf{Definition 3:(Bilinear CAA1 (k-BCAA1) [19])}. Let
 k be an integer, and x $\in$ ${Z_{q}}^{*}$, $P_{2}$ $\in$ ${G_{2}}^{*}$ ,
$P_{1}$ = $\psi$ $(P_{2})$, \^e : $G_{1} \times G_{2}$
$\longrightarrow$ $G_{T}$. Given ($P_{1}, P_{2}, xP_{2}, h_{0},
(h_{1}$, $\frac{1}{h_{1}+x}P_{2}), ... , (h_{k}$,
$\frac{1}{h_{k}+x}P_{2}$)), with $h_{i} \in Z_{q}$, for 0 $<$ i
$<$ k are distinct. Calculate \^e($P_{1}$,
$P_{2})^{\frac{1}{(x+h_{0})}}$
is difficult.\\
\textbf{Definition 4: (Bilinear Diffie-Hellman Problem BDHP [2])}.
Let $G_{1}$, $G_{2}$ two rings with prime order  q. Let \^e :
$G_{1} \times G_{2}$ $\longrightarrow$ $G_{T}$ be an application
admissible and bilinear and let P be a generator of $G_{1}$. The
BDHP in $<$ $G_{1},G_{2}$, \^e $>$ is so: Given $<$ P, aP, bP, cP
$>$ for a, b, c $\in$ $Z_{q}$. Calculate \^e(P, $P)^{abc}$ $\in
G_{2}$ is
difficult.\\
\textbf{Definition 5:(Augmented Bilinear Diffie-Hellman Exponent
Assumption q-ABDHP [6])}. Let
 k be an integer, and x $\in$ $Z_{q}^{*}$,
$ P_{2} \in G_{2}^{*}$, $P_{1}$ = $\psi(P_{2})$, \^e : $G_{1}
\times G_{2}$ $\longrightarrow$ $G_{T}$, given $(P_{1},
x^{k+2}P_{1}, P_{2}, xP_{2}, x^{2}P_{2},..., x^{2k}P_{2})$.
Calculate \^e($P_{1},P_{2})^{x^{k+1}}$ is difficult.\\
  \textbf{Definition 6: Problem calculator of Diffie Hellman:
  CDHP}.
Given P, aP, bP can we find or rather calculate abP?\\
\textbf{Definition 7: Problem Decisional of Diffie Hellman}\\
 Given P, aP, bP, cP can we say that
abP = cP?. But this problem can be solved in polynomial time after
using the pairing, for example if we prove that: e(P,cP) =
e(aP,bP) so abP = cP. This strategy is valid to others problems
for example the q-BDHIP and q-ABDHE
\subsubsection*{2.1.1-2 Relation} Firstly, we discuss and show the
relationship between the problems of Bilinear Diffie Hellman, with
which the studies of simulations of the cryptosystems in
competition are based. Study the classification of these problems
is useful, because the rigidity of these studies  is based on
them. So we have:
\begin{center}
BDHP (1) $\longrightarrow$ BDHIP (2)\\
BDHP (1) $\longrightarrow$ ABDHP (2)\\
BDHP (1) $\longrightarrow$ DBDHP (3)\\
BDHIP (2) $\longrightarrow$ DBDHIP (4)\\
ABDHP (2) $\longrightarrow$ DABDHP (4) \subsubsection*{Relation
and Classification}\end{center} We have classed DBDHP in class 3
compared with BDHIP and ABDHP, because, it can be calculated in
polynomial time using the Pairing. And we give the same rank to
ABDHP and BDHIP, since until present, there is no relationship
which can link these two problems, all we can say is that they
belong to the same category
(queries in the form exponentiations).\\
As long as, DBDHP has a rank before that of DABDHP and DBDHIP,
 because, (ignoring that  BDHP $\longrightarrow$ ABDHP and
BDHIP) the BDHP is rigid than BDHIP and ABDHP. Since theses latter
have complexity O($\sqrt[3]{q}$) after [20]. So, the DBDHP is also
rigid than DABDHP and DBDHIP. Recall that: BF (BDHP), SK (BDHIP),
BB1 (BDHP) BB2
(DBDHIP), Water (DBDHP), Gentry (DABDHP).\\
In the other part,  IBE has been built to serve a broad category
of a persons (in a classified area), using a single system of
parameter. The only things that is change is the private keys,
which are generated from a single master key for all the
applications. So it may be that there exist enemies among the
customers (the domains), who are agree to break the  Master key of
the authority from the syntax of the private key. So the success
of this study related  to the syntax of each  private key.\\ The
private key of the cryptosystems in competition are in the form:
BF has the form SiE BDHP ($sQ_{ID_{i}}$ for each i varied), that's
of SK has the form BCAA1 ($\frac{1}{s+H(_{ID_{i}})}$). BB1 is
based on PDL, as so not to extract
 $\alpha $, $\beta$, $\varpi$  from
respectively $\alpha P_{pub}$, $\beta P_{pub}$, $\omega P_{pub}$.
Also, we wouldn't calculate  $P_{prive}$ from  $rP_{prive}$,
since, if this will be easy, it will be easy also to associate a
random r to($\alpha H(ID)+\beta) P_{prive} +\omega P_{prive}$. So,
breaking easily the cryptosystem as we have the division of two
Pairing. For BB2 it has the private key following the form BCAA1
($\frac{1}{s_{1}+ID_{i}+s_{2}r)}$). The syntax of the private key
of Water is like BB1
based on PDL, as that of  Gentry is under the form BCAA1.\\
As it is generally known the PDL has complexity O($\sqrt{q}$) and
the BCAA1 has  O($\sqrt[3]{q}$) [20], as it is from the category
of the Problem Diffie Hellman in form Exponentiations. For the
SiE-BDHP we haven't a  complexity exact, all we can say is that it
is less than PDL and more than BCAA1, since PDL $\longrightarrow$
SiE-BDHP $\longrightarrow$ EBDHP $\longrightarrow$ BCAA1  (EBDHP
Exponent Bilinear Diffie Hellman Problem [19]). So we have this
classification  following the rigidity of the private key : BF(2),
SK(3), BB1(1), BB2(3), Water(1), Gentry(3) \subsubsection{Random
Oracle \& Standard Model} Random Oracle : In cryptography, an
oracle is a random that answers all queries proposed at random and
specific request (for
more details we send the interested to[21])\\
The utilization of the  Random Oracle has some dangers,  we
cite in this article:\\
The Random Oracle responds with random values and therefore, it
will be difficult to precise the suitability of its values with
the conditions allowed. More, because of the random values of the
Random Oracles which are difficult to adapt, the crypto systems
under this model use in their demonstrations  an  arbitrarily
values chosen. Which makes these cryptosystems unclear in their
study of simulations ($q_{H}$ is not related directly to the
syntax of the cryptosystem  but it is arbitrary). The Random
Oracle still has more danger and to knowing it we refer the
interested to [22]. By contrast, in  the Standard Model, which use
any Random Model we are sure about what is happening, as we use
the Mathematical formulas. But in the Random Oracle we communicate
with a spirit random which hasn't any exact measure.
 \subsubsection{Studies of Simulations}
The studies of simulations are invented by [12], they are being
done in advance to test the rigidity of a cryptosystem. And in
this article we cite :
\\ \textbf{ CPA :} Is the
abbreviation of Chosen Plaintext Attack ie during the studies of
simulations the opponent has advantage to
access to the encrypted of his chosen texts.\\
\textbf{ CCA :} It is an abbreviated of Chosen Ciphertext Attack,
and we divide it into two parts: CCA1 and CCA2.
During  CCA the adversary has advantage of access to the decrypts
texts he has chosen. In the CCA1 the opponent is less limited by
comparison with CCA2. We must say that the CCA2 is the most
powerful among all these attacks.\\
In 2003 Canetti, Halevi and Katz proposed an alternative strategy
in the study of simulation,  at which the adversary must commit
ahead of time to the challenge identity. And so, the identity to
attack must be  declared in advance. This early model is referred
as selective-identity attack (sID), while the Original Model is
called Full-identity scenario (ID). According to [23] the
selective ID (sID-CCA/CCP) is less rigid than (ID-CCA/CCP). The
ID-CCA  is required to merit the Standardization.
\subsubsection{Advantage of the Cryptosystem} In this section, we
compare the advantage of each cryptosystem in competition. Recall
that an advantage is done to learn the skill of an opponent to
break a cryptosystem, basing on
specifically mathematical probabilities. For our cryptosystems we have:\\
$Adv_{BF}$ (Advantage of BF) =
$\frac{1}{(q_{H_{3}}+q_{H_{4}})q_{H_{2}}}[(\frac{\varepsilon}{q_{H_{2}}}
  (1-\frac{q_{E}}{q_{H_{1}}})+1)(1-\frac{2}{p})^{q_{D}}-1]$ $-\frac{3}{6}$
  $\sim$
  $\frac{\varepsilon}{q_{H^{3}}}$$(1-\frac{2}{p})^{q_{D}}$;
 $Adv_{SK}$ =
$(\frac{\varepsilon}{q_{1}+1})
  (1-\frac{2}{p})^{q_{D}}$.
For the two crypto system BB1 and BB2 we utilize a propriety
demonstrated by Boneh Boyen [4] which say that: \\Let
 (t, $q_{S}$, $\varepsilon$)-selective
identity secure IBE system (IND-sID-CPA). Suppose E admits N
distincts identities. Then E is also a (t, $q_{S},N
\varepsilon$)-fully secure IBE (IND-ID-CPA). So basing in this
propriety we have:
  $Adv_{BB1}$  = $\varepsilon.2^{n}.\frac{q_{H}}{(2^{n} -
  q_{S})}$;
 $Adv_{BB2}$  = $\varepsilon.2^{n}$.
As long as following [5] and [6] we extract easily: $Adv_{Water}$=
$\frac{\varepsilon}{32(n+1)q}$; $Adv_{Gentry}$= $\varepsilon +
4\frac{qC}{p}$. To compare this advantages  we take into
consideration:\\ $q_{S}$ \&  $q_{D}$ $<$ $q_{H}$ $<$ n $<$$<$ p.
So we have:\\
$Adv_{Water}$ $<$ $Adv_{BF}$ $<$ $Adv_{SK}$ $<$ $Adv_{Gentry}$ $<$
 $Adv_{BB1}$ $<$ $Adv_{BB2}$. Consequently, Water is  the most desirable
as it has a very small advantage  \vspace*{-.08cm}
\subsubsection{Anonymity} Anonymity is a method to distinguish the
identity of a person from the ciphertext. This property is more
desirable in cryptography, because it limits the activity of an
opponent in the beginning. As a result, the opponent will  be
incapable to know the person addressed in the ciphertext. For our
cryptosystems only Boneh Franklin and Gentry are Anonymous
 \subsubsection{Pairing}
 A pairing is a bilinear map that takes two points on
an elliptic curve and gives an element of the group multiplicative
of n-th roots of unity.
Among the pairing we cited: Weil, Tate, Ate, $\eta$, but in the
implementations cryptographic we often use Weil and Tate.
\subsubsection*{Pairing of Weil}
 The Weil pairing is defended as follows: $e_{r}$:$E[r] \times
E[r] \rightarrow \mu_{r}$ ($\mu_{r}$ is the set of the $r^{th}$
root of the unity) such that:
$e_{r}(P,Q)=\frac{f_{D_{Q}}(D_{P})}{f_{D_{P}}(D_{Q})}$
\subsubsection*{Pairing of Tate}
The Tate pairing is the application :\\
$t_{r}$:E(k)[r]$\times$E(k)/rE(k) $\rightarrow$
$k^{\ast}/(k^{\ast})^{r}$\\
(P,Q)$ \rightarrow$$t_{r}(P,Q)$=$f_{D_{P}}(D_{Q})$ modulo
$(k^{\ast})^{r}$.
And to have an exact value, it can be defined as follows:
$t_{r}(P,Q)= (f_{D_{P}}(D_{Q}))^{(q^{k}-1)/r}$
\subsubsection{Inverse of two Pairing}
The inverse of two pairing is  calculate as [24]\\
$\frac{e(P_{1}, Q_{1})}{e (P_{2},Q_{2} )} = e (P_{1},Q_{1})e
(P_{2},-Q_{2} )$, and if we take $P_{1}$ and $P_{2}$ with the same
order,  we can so utilize the same algorithm of Miller to
calculate the inverse of two pairing. The only things we change is
instead of $f_{1}$ $\leftarrow$ ${f_{1}}^{2}$ $\times$
$\frac{l_{1}(Q_{1})}{v_{1}(Q_{1})}$ we calculate  $f_{1}$
$\leftarrow$ ${f_{1}}^{2}$ $\times$
$\frac{l_{1}(Q_{1})}{v_{1}(Q_{1})}$$\times$
$\frac{l_{1}(Q_{2})}{v_{1}(Q_{2})}$  also instead of\\$f_{1}$
$\leftarrow$ $f_{1}$ $\times$ $\frac{l_{2}(Q_{1})}{v_{2}(Q_{1})}$
we calculate $f_{1}$ $\leftarrow$ $f_{1}$ $\times$
$\frac{l_{2}(Q_{1})}{v_{2}(Q_{1})}$ $\times$
$\frac{l_{2}(Q_{2})}{v_{2}(Q_{2})}$.\\\\
The calculation of the pairing is ineffective until the invention
of the algorithm of Miller in 1986.\\\\\\\\\\
\begin{table}[h!]
\begin {tabular}{|c|}
\hline
\textbf{Miller(P, Q, r)}\\
\hline\\
 \hspace{+1.2cm}\textbf{Input: r =
($r_{n}$...$r_{0}$)(binary representation )},
\\ \hspace{.75cm} P $\in$ $E[r]$($\subset$ E($F_{q}$)) and
Q $\in$ $G_{1}$($\subset$ $E(F_{q^{k}}$)) \\
 \hspace{-2.1cm}\textbf{Output:} $f_{r,P}(Q)$ $\in$
$G_{3}$ ($\subset$ $F_{q^{k}}^{*})$
\\ \hspace{-5.1cm} T $\leftarrow$ P\\
 \hspace{-5.1cm}  $f_{1}$ $\leftarrow$ 1 \\
 \hspace{-3.5cm} for i = n - 1 to 0 do \\
 \hspace{-4.5cm} \textbf{1:} T $\leftarrow $  [2]T \\
 \hspace{-2.84cm}  $f_{1}$ $\leftarrow$
${f_{1}}^{2}$ $\times$ $\frac{l_{1}(Q)}{v_{1}(Q)}$ \\
\hspace{-1.2cm} $l_{1}$ is the tangent to the curve in T.\\
\hspace{-.8cm} $V_{1}$ is the vertical to the curve in [2]T.\\
 \hspace{-4.cm} \textbf{2:} if $r_{i}$=1 then \\
 \hspace{-2.94cm}  $f_{1}$ $\leftarrow$
$f_{1}$ $\times$ $\frac{l_{2}(Q)}{v_{2}(Q)}$\\
\hspace*{-.4cm} $l_{2}$ is the line  passing through the point TP\\
\hspace*{-1.4cm} $V_{2}$ is the vertical to the point P + T.\\
\hspace{-3.5cm}\textbf{Output: Return $f_{1}$ }\\
\hline
\end{tabular}
\end{table}\\
 \normalsize
 \subsubsection{Haching on an elliptic Curve}
In the cryptosystem of Boneh and Franklin  there is, the problem
of Hashing Function  in an elliptic curve selected. And to do it
we remember the  method suited by Boneh Franklin
\subsubsection*{Map to point} \hspace*{-.02cm}0. Project  the ID
using: $H_{1}$ : ID
$\in$ $\{0,1\}^{*}$ $\longrightarrow$ $y_{0}$ $\in$ $F_{p}$    \\
1. Calculate $x_{0}$ = $({y_{0}}^{2}-1)^{\frac{1}{3}}$ =
$({y_{0}}^{2}-1)^{\frac{2p-1}{3}}$ $\in F_{p}$. \\ 2. Let Q =
$(x_{0},y_{0}) \in E(F_{p})$ after calculate  $Q_{ID} = lQ \in G.$
\\3. Output MapToPoint($y_{0}) = Q_{ID}$.
\subsubsection{Cryptosystems in Competition} The cryptosytems in
competition are Boneh and Franklin, Skai Kasarah, Boneh Boyen,
Water, Gentry.  In this article we choose them, taking into
account the most recent changes to make them effective. So for
Boneh and Franklin we prefer to use that of Galnido [25] instead
of the version of Boneh and Franklin. Because Galnido provide
reduction in the advantage of Boneh Franklin. More, this latter is
valid only on supersingular curve, as it uses symmetric pairing.
By contrast, Galnido use asymmetric pairing of type II and he
established his argument based on them. Following [26] the
asymmetric pairing, with which we can use ordinary curves are more
convenient in implementations than the symmetric one.
\vspace*{-.1cm}
\begin{table}[h!]
\begin{tabular}{|c|}
  \hline
  \large Boneh-Franklin (Galindo-Full Version)\\
  \hline
\hspace*{-.6cm}\textbf{Setup}. Let ($G_{1}, G_{2}, G_{T} ,\psi$) a
bilinear group. Choose a generator \\ \hspace*{-.4cm}$P_{2}\in
G_{2}$ and set $P_{1} = \psi(P_{2})$. Next pick s$ \longleftarrow$
$Z_{p}$ \\\hspace*{-.45cm} and set $Q_{pub} = sP_{2} \in
{G_{2}}^{\star}$ $\rightarrow$ $P_{pub} = sP_{1} \in {G_{1}}^{*}$.\\
\hspace*{+1.5cm} Choose cryptographic hash functions $H_{1} :
{\{0, 1\}}^{*}$ $\longleftarrow {G_{2}}^{\star}$ ,\\
\hspace*{+.2cm} $H_{2} : G_{T} \longleftarrow \{0, 1\}^{n}$,
$H_{3} : \{0, 1\}^{n} \times \{0, 1\}^{n}$ $\longleftarrow$
${Z_{p}}^{*},$\\ \hspace*{+1.3cm} $H_{4} : \{0, 1\}^{n}$
$\longleftarrow \{0, 1\}^{n}$. The message space is M= $\{0,
1\}^{n}$  \\ \hspace*{+1cm} and the ciphertext space is C
= ${G_{1}}^{*} \times \{0, 1\}^{n} \times \{0, 1\}^{n}$. \\
\hspace*{-.2cm} \textbf{Extract}. For a given string ID $\in \{0,
1\}^{*}$, compute $Q_{ID} = H_{1}(ID)$\\
\hspace*{+.9cm} and set the private key $d_{ID}$ to be $d_{ID} =
sQ_{ID} \in {G_{2}}^{*}$.
\\ \hspace*{-1.1cm}\textbf{Encrypt}. To encrypt M $\in {0, 1}^{n}$ under identity ID,
compute \\ \hspace*{-.2cm} $Q_{ID }$=  $H_{1}(ID) \in
{G_{2}}^{*}$, choose $\sigma \longleftarrow \{0, 1\}^{n}$,\\
\hspace*{-1.7cm}set r = $H_{3}(\sigma,M) \in {Z_{p}}^{\star}$ and
finally \\\hspace*{-.7cm}C = $<$ $rP_{1}, \sigma \bigoplus
H_{2}(g_{ID}^{r}), M \bigoplus H_{4}(\sigma)$ $>$ \\
\hspace*{-2cm}where $g_{ID} = e(P_{pub},Q_{ID})\in G_{T}$.
\\ \hspace*{+.4cm}\textbf{Decrypt}. Let C = $<$ U, V,W $>$  $\in$ $C$ be a
ciphertext under the identity \\\hspace*{+.9cm} ID. To
decrypt C using the private key $d_{ID} \in {G_{2}}^{\star}$ do:\\
\hspace*{-1.6cm} 1. Compute $ V \bigoplus H_{2}(e(U, d_{ID})) =
\sigma$.\\ \hspace*{-2.5cm}2. Compute $ W \bigoplus H_{4}(\sigma)
= M$.\\ \hspace*{-0.6cm}3. Set r = $H_{3}(\sigma,M)$. Check that U
= rP. \\ \hspace*{-2.2cm} If not, reject the ciphertext.\\
\hspace*{-5.6cm} 4. Output M.\\\hline
\end{tabular}
\end{table}\\
\begin{table}[h!]
\begin{tabular}{|c|}
  \hline
   Sakai-Kasaharah (ChenCheng-Full Version)\\
  \hline
 \hspace*{-.6cm}\textbf{Setup}. Let ($G_{1}, G_{2}, G_{T} ,\psi$)
a bilinear group. Choose a generator \\ \hspace*{-1.2cm}$P_{2}\in
G_{2}$ and set $P_{1} = \psi(P_{2})$. Next pick s$ \longleftarrow$
$Z_{p}$ \\\hspace*{-.1cm} and set $Q_{pub} = sP_{2} \in
{G_{2}}^{\star}$ $\rightarrow$ $P_{pub} = sP_{1} \in {G_{1}}^{*}$.
Choose \\ \hspace*{-1.cm}crypto graphic hash functions $H_{1} :
{{0, 1}}^{*}$ $\longleftarrow {G_{2}}^{\star}$, \\
\hspace*{-.4cm}$H_{2} : G_{T} \longleftarrow \{0, 1\}^{n}$, $H_{3}
: \{0, 1\}^{n} \times \{0, 1\}^{n}$ $\longleftarrow$
${Z_{p}}^{*}$,
\\
\hspace*{+.5cm} $H_{4} : \{0, 1\}^{n}$ $\longleftarrow \{0, 1\}^{n}$. The message space is M= $\{0, 1\}^{n}$  \\
\hspace*{+.2cm}and the ciphertext space is C = ${G_{1}}^{*} \times
\{0, 1\}^{n} \times \{0, 1\}^{n}$.
\\\hspace*{-.2cm} \textbf{ Extract:} Given an identifer string $ID_{A}$ $\in$
$\{0,1\}^{n}$  of entity A, $M_{pk}$ \\ \hspace*{+.1cm} and
$M_{sk}$,
 the algorithm returns
$d_{A}$=$\frac{1}{s+H_{1}(ID_{A})}$ $P_{2}$\\
\hspace*{-2.2cm}\textbf{ Encrypt:} Given a plaintext m $\in$ M ,
$ID_{A}$ and $M_{pk}$, \\ \hspace*{-3.3cm}the following step are
formed:\\\hspace*{+.4cm} 1.pick a random $\sigma$ $\in$
$\{0,1\}^{n}$ and compute
r=$H_{3}(\sigma,m)$\\
\hspace*{+.1cm} 2.Compute $Q_{A}=H_{1}(ID_{A})P1+P_{pub}$,
$g^{r}$=$e(P_{1},P_{2})^{r}$ \\
\hspace*{+1.5cm}Set the ciphertext to be $C=(rQ_{A},\sigma \oplus
H_{2}(g^{r})$, m
$\oplus H_{4}(\sigma))$\\
\hspace*{-1.6cm} \textbf{Decrypt:} Given a ciphertext C = (U,V,W)$\in$$C$, $ID_{A}$, $d_{A}$ \\
\hspace*{-4.6cm} and $M_{pk}$, follow the steps\\
\hspace*{-1.2cm} 1.Compute g'=e$(U,d_{A})$ and $\sigma'=V \oplus H_{2}(g')$\\
\hspace*{-.7cm} 2.Compute m'=W $\oplus$ $H_{4}(\sigma')$ and r'=
$H_{3}(\sigma',m')$\\\hspace*{-1.6cm}3.If U $\neq$
$r'(H_{1}(ID_{A})P_{1}+P_{pub})$ output $\perp$
\\\hspace*{-3.cm} else
return the m' as the plintext\\
  \hline
\end{tabular}
\end{table}\\\\\\\\\\\
\begin{table}[h!]
\begin{tabular}{|c|}
\hline
\Large Boneh-Boyen\\ \hline \normalsize BB1(Full Version)
  \\
  \hline
 \hspace*{-.8cm}  \textbf{Setup:} \small To generate IBE system parameters,
 pick $\omega, \alpha,$ \\\hspace*{+.5cm} $\beta, \gamma \in Z_{p}$, and
   output,
params = \{ P, $P_{1} =  \alpha P,$ \\ \hspace*{-1.2cm} $P_{2} =
\beta P,$, $v_{0} = e(P,\hat{P})^{\omega} \} \in {G_{1}}^{3} \times G_{t}$,\\
\hspace*{-1.8cm} masterk = $(\hat{P},\omega,\alpha,\beta)\in G_{2} \times {Z_{p}}^{4}$.\\
  \hspace*{+.8cm} Let $g_{1}$ and $g_{2}$ be the respective
  generators of some \\ \hspace*{+1.5cm}bilinear group pair $(G_{1},G_{2})$ of prime order p,
  And let  \\ \hspace*{+.6cm}e : $G_{1}$ $\times$ $G_{2}$ $\longrightarrow$
  $G_{t}$ be a bilinear pairing map.
  \\ \hspace*{1.4cm} The availability of
three cryptographic hash functions \\
\hspace*{+.6cm} viewed as  random oracles
  graphic hash functions \\\hspace*{-.55cm}$H_{1} $:
  ${\{0,1\}}^{*}$ $\longleftarrow$ $Z_{p}$
, $H_{2} : G_{t} \longleftarrow \{0, 1\}^{n}$,\\
\hspace*{-1.5cm} $H_{3} : G_{t}\times \{0, 1\}^{n} \times G_{1}
\times$ $G_{2}$ $\longleftarrow $ $Z_{p}$.\\ \hspace*{-1.3cm} The
message space is M= $\{0, 1\}^{n}$ and \\ \hspace*{+1.1cm} The
ciphertext space is
C = ${G_{1}}^{*} \times \{0, 1\}^{n} \times \{0, 1\}^{n}$.\\
\hspace*{-.5cm}\textbf{Extract:} To extract from masterk a private
key $d_{ID}$ for an \\ \hspace*{+2cm} identity ID$\in$ $\{0,
1\}^{l}$ , pick
a random r$\in$ $Z_{p}$ and output\\
\hspace*{+1.3cm}$d_{ID} = (d_{0} = (\omega + (\alpha H_{1}(ID) +
\beta])r)\hat{P}$, $d_{1}$ = $r \hat{P} $).\\
\hspace*{-1.2cm}\textbf{Encrypt:} Given a plaintext m $\in$ M ,
$ID_{A}$
and $M_{pk}$,\\ \hspace*{-1.4cm} the following step are formed:\\
\hspace*{.85cm}$C$ = $\left\{%
\begin{array}{ll} & c = \hbox {M $ \bigoplus $  $H_{2}(k = v_{0}^{s})$,}\\ & \hbox {$c_{0}$ = sP,}\\
& \hbox {$c_{1} = H_{1}(ID) sP_{1} + sP_{2}$,}\\ & \hbox{ t = s +
$H_{3}(k, c, c_{0},
c_{1})$ mod p )}\end{array}%
\right.$\\
\hspace*{+.8cm}where M $\in$ \{0,1\} is the message, ID $\in$ \{0, 1\}\\
\hspace*{-.6cm} is the recipient identifier, and
$s \in Z_{p}$\\ \hspace*{-1.5cm} is a random ephemeral integer.\\
\hspace*{+.3cm} \textbf{ Decrypt:} Given a ciphertext C and a
private key $d_{ID} = (d_{0}, d_{1})$,\\ \hspace*{+.2cm} compute,
k = $\frac{e(c_{0},
d_{0})}{e(c_{1}, d_{1})}$,  s= t - $H_{3}(k, c, c_{0}, c_{1})$.\\
\hspace*{-1cm}If $(k, c_{0}) \neq$ = (
$v_{0}^{s}$, sP ), output $\perp$; \\ \hspace*{-1cm} otherwise, output, M = c $\bigoplus$ $H_{2}(k)$.\\
\hline
\normalsize BB2 (Version CPA)
\\
\hline \hspace*{-4cm} \textbf{Setup} outputs Msk $\longleftarrow$
(a,b) and \\ \hspace*{-.8cm}Pub $\longleftarrow$ ( P, $P_{a}$ =
aP, $P_{b}$ = bP, v = $e(P,\hat{P}))$
\\\hspace*{-3.cm}for a, b $\in$
$F_{p}$ chosen at random.\\
\hspace*{-5cm} \textbf{Extract}(Msk,Id) outputs
\\\hspace*{+.1cm}$Pvk_{Id}$ $\longleftarrow$ ( $r_{Id} = r, .
\hat{d_{Id}} = \frac{-1}{ a+Id+b r} \hat{P} )$ for
$r \in F_{p}$\\
\hspace*{-3.9cm}\textbf{Encrypt}(Pub, Id, Msg, s) outputs
\\\hspace*{+.3cm}Ctx $\longleftarrow$
$(c_{0} = Msg.v^{s}, c_{1} = sP_{a} + sIdP, c_{2} = sP_{b}).$\\
\hspace*{-3.5cm}\textbf{Decrypt}(Pub, $Pvk_{Id}$, Ctx) outputs
\\\hspace*{-2cm} Msg' $\longleftarrow$ $c_{0}.e(c_{1} + r_{Id}c_{2},
\hat{d_{Id}}) \in G_{t}.$ \\\hline
\end{tabular}
\end{table}\\
\begin{table}[h!]
\begin{tabular}{|c|}
  \hline
  \Large  Water (Naccache-Version CPA)\\
  \hline
   \textbf{Setup:}\small Choose a secret parameters $\alpha$ $\in$ $Z_{p}$
  at random,\\ \hspace*{+.6cm} choose a random generator g $\in$ G  and set the value
  \\ \hspace*{-1cm} $g_{1}$ = $\alpha$g  also choose at randomly $g_{2}$ $\in$ G.
  \\ \hspace*{-.3cm} The authority choose a random value u' $\in$ G \\\hspace*{+.2cm} and a random n
  length vector U=($u_{i}$) chosen at \\  \hspace*{-.35cm}random from G. The
  publish parameters are \\ \hspace*{+.3cm}  params $<$ g,$g_{1}$,$g_{2}$,u',U $>$ the
  master secret is $\alpha$$g_{2}$
  \\ \hspace*{-.6cm} \textbf{Key Generation}\small : Let v = $(v_{1},..., v_{n}) \in (\{0,1\}^{a})^{n}$\\
  \hspace*{-1.6cm}be an identity,
  Let r be random in $Z_{p}$\\ \hspace*{-.8cm} The private key $d_{v}$ for identity v is
  construc\\
  \hspace*{-1.4cm}ted as :
  $d_{v}$ = $(\alpha g_{2}+ r (u'+\sum_{i=1}^{n}u_{i})$,rg)\\
 \hspace*{-1.5cm}  \textbf{Encryption:}\small A message M $\in$ $G_{1}$ is
  encrypted \\\hspace*{-3cm}  for an identity v as follows.\\ \hspace*{-1.4cm}A value t $\in$
  $Z_{p}$ is chosen at random\\ \hspace*{-1.1cm}
  The ciphertext is then constructed as:\\
 \hspace*{-1.2cm}  C=($e(g_{1},g_{2})^{t}$M, t.g, t.$(u'+\sum_{i=1}^{n}u_{i})$))\\
 \hspace*{-.5cm} \textbf{Decryption:}\small Let C=$(c_{1},c_{2},c_{3})$ be a
valid encryption\\ \hspace*{-.6cm} of M under the identity v. Then
C can be
\\\hspace*{-.8cm} decrypts by $d_{v}$=$(d_{1},d_{2})$ as:
$c_{1}$$\frac{e(d_{2},C_{3})}{e(d_{1},C_{2})}$=M\\
  \hline
  \end{tabular}
\end{table}
\begin{table}[h!]
\begin{tabular}{|c|}
\hline
 \Large  Gentry(Full-Version) \\
  \hline
  \ \textbf{Setup:}\small The PKG picks a random generators  $<$$g,h_{1},h_{2},h_{3}$
  $>$\\
 \hspace*{+.55cm}and a random $\alpha$ $\in$ $Z_{p}$. It sets $g_{1}$ =
  $\alpha$g $\in$ G. It chooses a \\\hspace*{+.34cm} hash function H from a family
  of
  universal one-way hash \\\hspace*{+.5cm}functions. The public params and
  private master-key are \\\hspace*{-.08cm} given by
 params = $<$g,$g_{1}$,$h_{1}$,$h_{2}$,$h_{3}$,H$>$ master-key=$\alpha$\\
\hspace*{-.4cm} \textbf{Key Gen:}\small To generate a private key
for identity ID $\in$ $Z_{p}$,\\\hspace*{-1.4cm}the PKG generates
random $r_{ID,i}$ $\in$ $Z_{p}$ for \\\hspace*{-2cm} i $\in$
\{1,2,3\} and output the private key\\\hspace*{-2cm}
$d_{ID}$=\{($r_{ID,i}$,$h_{ID,i}$: i $\in$ \{1,2,3\}, where
\\\hspace*{+.8cm} $h_{ID,i}$=$\frac{1}{\alpha-ID}$($h_{i}$+($r_{ID,i}$g))
If ID = $\alpha$, the PKG aborts.\\
\hspace*{-.5cm} \textbf{Encrypt:}\small To encrypt m $\in$ $G_{T}$
using identity ID $\in$ $Z_{p}$, the \\\hspace*{-.2cm}sender
generates random s $\in$ $Z_{p}$ and send  the \\\hspace*{-.5cm}
ciphertext
$C$ = $\left\{%
\begin{array}{ll} & \hbox {u=$sg_{1}+(-sID)g$,}\\
& \hbox {v=$e(g,g)^{s}$,}\\
& \hbox {w=$m.e(g,h_{1})^{-s}$,}\\ & \hbox{ y=$e(g,h_{2})^{s}$$e(g,h_{3})^{s\beta}$}\end{array}%
\right.$\\\hspace*{-.55cm}
Above, for C=(u,v,w,y) we set $\beta$=H(u,v,w)\\
\hspace*{-1cm} \textbf{Decrypt:}\small To decrypt ciphertext
C=(u,v,w,y)
with ID \\ \hspace*{-.1cm} the recepient sets $\beta$=H(u,v,w) and test wether \\
\hspace*{-.8cm}
y=e(u,$h_{ID,2}h_{ID,3}^{\beta}$)$v^{r_{ID,2}+r_{ID,3}\beta}$ If
the check \\\hspace*{.49cm} fails, the recipient output $\perp$.
Otherwise, it
outputs\\ \hspace*{-4.cm}m=w.$e(u,h_{ID,1})v^{r_{ID,1}}$\\
\hline
\end{tabular}
\end{table}\\\\\\\\
\subsubsection*{Justification of the Choose}
\vspace*{-.2cm} We are making our choose based on the recent
modifications concerning the cryptosystems in competition. For
that of Boneh and Franklin, we have justified the version of
Galnido. As that of Skai Kasarah, we prefere to use the version of
Chen-Cheng [19] which is CCA secure. As far as concerned, the
version of BB1 we will  utilize the Random oracle version, such
that BB1 has a lot of versions: Random Oracle, selectiveID, and
also Standard Model. We will only play on the $H_{1}$, but we
prefer the  first one, because we have the cryptosystem of  Water
which has the same syntax as BB1 and is under  Standard Model. As
long as, that of Water we will use the version of Nackache which
utilize the Words instead of the alphabet. And  this reduce the
complexity \vspace*{-.3cm}
\subsection{Efficient Comparison} \vspace*{-.1cm} As we
have signaled Xavier. Boyen in 2008  essayed to make the
comparison [11] between Boneh Franklin, Skai Kasarah and BB1.  By
counting for example the numbers of the parameters for each
cryptosystem, the groups associates, the propriety associates.
More he has calling to the standardization of the cryptosystem BB1
[10] using the same method. Unfortunately his essay isn't
practical for the raison that he don't compute the complexity
exact (spatial and temporal) for each cryptosystem. He fixed only
the basis and he bagun to compute following the number of the
parameters. He posed some critters and he verified if only the
cryptosystems has it or not without demonstrate any
classification. By contrast, in our comparison we will follow
another strategy. We  pose a scale which we make in the
consideration the utility of the propriety,  this allow us to
precise the best cryptosystem. \vspace*{-.3cm}
\subsubsection{Comparison in the level Security} Before staring
the comparison in the level of
security we remember firstly the following things:\\
\begin{table}[h!]
\begin{tabular}{|c|c|c|c|c|c|}
  \hline
    BF & SK & BB1 & BB2 & Water & Gentry  \\
  \hline
    RO &  RO &   RO \& sID &    RO \& sID &  SM &  SM \\
  \hline
   BDHP & BDHIP & BDHP  &  DBDHIP & DBDHP & Dq-ABDHP \\
  \hline
   CCA &  CCA &  CPA  &   CPA &  CPA &  CCA\\
  \hline
   SiE-BDHP & BCAA1 & PDL  &  BCAA1 & PDL & BCAA1\\
  \hline
\end{tabular}
\end{table}\\
To rank the crypto systems in direction security, we give the
scale following the usefulness of each propriety. Concerning the
model utilized: RO is the worst case as long as SM is the better,
until sID is between them,  therefore: RO (rank 3), sID (rank 2),
SM (rank 1). But because of the very great dangers of RO [22] and
as we presented a few of them  in section 2.1.2 we double these
coefficients in the table below. In the other part, because of the
utility of the anonymity for the security, as  it can early  block
the activity of the opponent we reducing the rank to 0 for those
that have it and we give 2 to those they don't have it. For the
remaining criteria we follow the classification we done in the
section 2.1.1 ; 2.1.4\\\\\\\\\\\\\\\\\\
\begin{table}[h!]
\hspace*{-.3cm} \caption{classification in the level security}
\vspace*{+.5cm}
\begin{tabular}{|c|c|c|c|c|c|c|}
  \hline
   & BF & SK & BB1 & BB2 & Water & Gentry \\
  \hline
  Model &  6 &  6 &   4 &   4 &  2 &  2\\
  \hline
  $Pro_{DH}$ & 1 & 2 & 1 & 4 & 3 & 4 \\
  \hline
  Avd & 2 & 3 & 5 & 6 & 1 & 4\\
  \hline
  Simu &  0 &  0 &  1 &  1&  1 &  0\\
  \hline
  $Pro_{DH_{priv}}$ & 2 & 3 & 1 & 3 & 1 & 3\\
  \hline
  Ano & 0 & 2 & 2& 2 & 2& 0\\
    \hline
Sum & 11 & 16 & 14 & 20 & 10 & 13 \\
\hline
Class & $(2^{sd})$ & $(5^{th})$ & $(4^{th})$ & $(6^{th})$ & $(1^{st})$ & $(3^{th})$ \\
  \hline
\end{tabular}
\end{table}
 \subsubsection{Comparison in the level
Complexity} In [10][11] Xavier Boyen tried to establish a base,
from which he tried to compte the time for the crypto systems that
are affected. But we can say that his results are not accurate
enough, because, he doesn't take into account some operations such
as: inverse, multiplication etc. By contrast in our study we
compte the most possibles operations. More our complexity can
combine between spatial and temporal
\subsubsection*{Complexity associate} We assemble our own complexity in the  following tables.\\
With the fact that in table III we set the parameters, with a
manner to reduce more possibly the calculation, for example,
instead of placing $g=e(P_{1},P_{2})$(in SK cryptosystem) in the
Encrypt at which we will recalculate it each time, we publish it
among the
Params\\
In the table IV the following symbol significate:\\ C: Complexity;
$Mul_{sca}$: Multiplication Scalar; $Exp_{ffi}$: Exponentiation in
the finite field; $Inv_{ffi}$: Inversion in the finite field;
$Mul_{ffi}$: Multiplication in the finite field; pair: Pairing;
Inv of 2 pair: Inversion of two pairing \\
\begin{table}[h!]
\hspace*{-.4cm} \caption{Parameter Associate} \vspace*{+.3cm}
\begin{tabular}{|c|c|}
  \hline
  $BF_{Ga}$ & $SK_{CC}$   \\
  \hline
   \hline
  \hline
   $ sP_{1}$ & $ sP_{1}$; $g=e(P_{1},P_{2})$  \\  \hline
  \hline
 $Q_{ID}$ (map to point);s$Q_{ID}$ & $\frac{1}{s+H_{1}(ID)}P_{2}$
     \\
 \hline
 \hline
  u=r$P_{2}$; $e(P_{pub},Q_{ID})^{r}$  & Q = $H_{1}(ID)P_{1}+P_{pub};g^{r};u=rQ $  \\
 \hline
  \hline
 $ e(u,d_{ID})$  & $ e(u,d_{ID})$;r'$Q_{A}$    \\
  \hline
\end{tabular}\\
\begin{tabular}{|c|c|c|}
  \hline
   BB1 & BB2 &  $Water_{Na}$\\
  \hline
   \hline
  \hline
   $ \alpha P_{1};\beta
P_{2};e(P,\hat{P})$;$v_{0}$ & $aP_{1};bP_{2};e(P,\hat{P})$& $\alpha g_{1};v = e(g_{1},g_{2})$\\
\hline
  \hline
 \hspace*{-.3cm} ($\omega + r(\alpha H_{1}(ID) + \beta))\hat{P}$; r$\hat{P}$
  \hspace*{-.3cm}  & $\frac{1}{a+ID+br}\hat{P}$ & $\alpha g_{2}+r(U'+{\sum^{n}}_{i=1}U_{i})$;rg \hspace*{-.3cm}\\
 \hline
 \hline
  $v_{0}^{s};sP;H_{1}(ID)sP_{1};sP_{2}$  & m.$v^{s}; sP_{a}; sIdP; sP_{b} $& $v^{t}; tg; t(U'+{\sum_{i=1}}^{n}U_{i})$\\
 \hline
  \hline
   $\frac{e(c_{0},d_{0})}{e(c_{1},d_{1})}$;$v_{0}^{s}$; sP  & $c_{0}.e(c_{1} + r_{Id}c_{2}, \hat{d_{Id}})$ & $c_{1}.\frac{e(c_{3},d_{2})}{e(c_{2},d_{1})}$\\
  \hline
\end{tabular}
\end{table}\\
\begin{table}[h!]
\begin{tabular}{|c|}
  \hline
       Gentry \\
  \hline
  \hline
       $v_{0}=e(g,g); v_{1}=e(g,h_{1});
   v_{2}=e(g,h_{2}); v_{3}=e(g,h_{3})$ \\
  \hline
  \hline
     $\frac{1}{\alpha - ID}(h_{i}+r_{ID,i}g), i \in \{1,2,3\}$
     \\
      \hline
  \hline
       u;  $v_{0}^{s}$;
$m.v_{1}^{-s}$; $v_{2}^{s}.v_{3}^{s \beta}$\\
       \hline
\hline

  $ y=e(u,h_{ID,2}+\beta{h_{ID,3}}){v_{0}}^{r_{ID,2}+r_{ID,3}\beta} $;w.$e(u,h_{ID,1})v^{r_{ID,1}}$\\
       \hline
\end{tabular}
\end{table}\\
\begin{table}[h!]
\caption{Complexity associate}
\begin{tabular}{|c|c|}
  \hline
  $BF_{Ga}$ & $SK_{CC}$  \\
  \hline
   \hline
  \hline
   C($Mul_{sca}$) & C($Mul_{sca}$)+C(pair)   \\  \hline
  \hline
  C(map to point)+C($Mul_{sca}$) & C($Inv_{ffi}$)+ C($Mul_{sca}$)
     \\
 \hline
 \hline
C($Mul_{sca}$)+C(pair)+C($Exp_{ffi}$)  &
2C($Mul_{sca}$)+C($Exp_{ffi}$)

  \\
 \hline
  \hline
 C(pair)  & C(pair)+C($Mul_{sca}$)    \\
  \hline
\end{tabular}
\\
\begin{tabular}{|c|}
  \hline
   BB1 \\
  \hline
   \hline
  \hline
    2C($Mul_{sca}$)+C(pair)+C($Exp_{ffi}$) \\  \hline
  \hline
   2C($Mul_{ffi}$)+ 2C($Mul_{sca}$)
    \\
 \hline
 \hline
\hspace*{-.25cm} 3C($Mul_{sca}$)+C($Exp_{ffi}$)+C($Mul_{ffi}$)
\hspace*{-.27cm} \\
 \hline
  \hline
 \hspace*{-.2cm} C(Inv of 2 pair)+C($Exp_{ffi}$)+C($Mul_{sca}$) \hspace*{-.2cm}\\
  \hline
\end{tabular}
\begin{tabular}{|c|}
  \hline
    BB2\\
  \hline
   \hline
  \hline
      2C($Mul_{sca}$)+C(pair)\\  \hline
  \hline
    C($Inv_{ffi}$)+ C($Mul_{sca}$)+C($Mul_{ffi}$)\\
 \hline
 \hline
 \hspace*{-.26cm} 3 C($Mul_{sca}$)+C($Exp_{ffi}$)+2C($Mul_{ffi}$)\\
 \hline
  \hline
   C($Mul_{ffi}$)+C(pair)+C($Mul_{sca}$)\\
  \hline
\end{tabular}
\\
\begin{tabular}{|c|c|}
  \hline
     $Water_{Na}$ \\
  \hline
  \hline
   C($Mul_{sca}$)+C(pair)  \\
  \hline
  \hline
 4 C($Mul_{sca}$)
     \\
      \hline
  \hline
 3 C($Mul_{sca}$)+C($Exp_{ffi}$)+ C($Mul_{ffi}$)

\\
       \hline
\hline C($Mul_{ffi}$)+C(Inv of 2 pair)
\\
       \hline
\end{tabular}
\begin{tabular}{|c|}
  \hline
      Gentry \\
  \hline
  \hline
   4C(pair) \\
  \hline
  \hline
  3 C($Mul_{sca}$)+C($Inv_{ffi}$)
     \\
      \hline
  \hline
 2 C($Mul_{sca}$)+ 4C($Exp_{ffi}$)+C($Inv_{ffi}$)+2C($Mul_{ffi}$)
\\
\hline \hline
4C($Mul_{ffi}$)+2C(pair)+\\C($Mul_{sca}$)+2C($Exp_{ffi}$)
\\
\hline
\end{tabular}
\end{table}\\\\
 \textbf{Observation:}To
calculate the Multiplication Scalar we consider in this article
that the operation of  adding and doubling are equal so for\\
example: ($U'+{\sum_{i=1}}^{n}U_{i})$ is considered as one Scalar
Multiplication.\\
\subsubsection*{ Complexity Neighboring} In this section we begin to fix the
complexity  for each cryptosystem. We can say that they are a
complexity neighbor, since we do not take into account: addition,
subtraction, calculation of hashed functions... More we balance
between the complexity of square with that of multiplication. Our
method help us to have a nearest comparison between the
cryptosystem's in competition, because we will concentrate only on
the main arithmetic (operation used): multiplication, square,
exponentiation, scalar
multiplication in each cryptosystem.\\
Following [27] we have:
:\\
1. C(compute of $m \times n$) = $O((log n)^{2})$ \\
2. C(compute of gcd(m, n)) = C(compute of $m^{-1})= O((log
n)^{3})$ = C(compute of $m^{-1}$ (mod n)) = $O((log n)^{3})$\\
For the exponentiation  we consider in this article the algorithm
Right-to-left
binary exp [28] which has complexity equivalent to:\\
$(\frac{1}{2}$lgn)Mu + ($lgn$)Sq = $(\frac{3}{2}$ lgn)Mu (as
declared C(Mu)=C(Sq) ). Those complexity are not a persuade
complexity and to make an exact one we will use the newest method
used in the literature. But this help us to order the main
operation in arithmetic, as [29] we have according to those
complexity:
C(multiplication) $<$ C(inverse) $<$ C(exponentiation)\\
In [11] Boyen balance between exponentiation $x^{n}$ and the
scalar multiplication [n]P as we can apply the same operations to
crush the n. This is not true,
because we must consider for [n]P an additional complexity:\\
Following [29], in jacobian coordinate we have:\\
C(ECADD)=12Mu+2Sq=14$O((log n)^{2})$ (C(Mu)=C(Sq) the Z $\neq$ 1)
\\And
C(ECDBL)=7Mu+5Sq=13$O((log n)^{2})$ (a $\neq$ -3)\\
With ECADD: designs elliptic curve point adding P+Q, ECDBL :
designs elliptic curve point doubling 2P.\\ Also following [29]
and using NAF algorithm we have:\\
C(dP)=(n-1)ECDBL+$\frac{(n-1)}{3}$ECADD=13(n-1)$O((log
n)^{2})$+14$\frac{(n-1)}{3}O((log
n)^{2})$=$\frac{53}{3}$(n-1)$O((log n)^{2})$.
\\And
C($2^{n}$P)=4nMu+(4n+2)Sq=(8n+2)$O((log n)^{2})$ i.e for d=$2^{n}$.\\
According to algorithm Maptopoint we have:\\ C(Maptopoint)= C(1
square) + C(1 cubic root) + C(1 multiplication scalar)\\ So:
C(Maptopoint) = $O((log n)^{2})$ + $O(lg lg n)$ +
$\frac{53}{3}$(n-1)$O((log n)^{2})$
 (complexity of the cubic root  is $O(lg lg n)$ following an algorithm in [28])\\
For the complexity of the pairing we will  take into
consideration, as possible all the reduction we can apply to
reduce the pairing. We take for example Tate because Weil
is heavy (two time bigger than Tate). So we have:\\
C(pairing=Tate)=C(Miler)+C(Exponentiation), since $t_{r}=(f_{r})^{\frac{q^{k}-1}{r}}$\\
With a naive calculate we have: \\Starting   with the complexity
of  the algorithm of Miller. We neglect  as customary to
accelerate  the compute, the second tranche of  the algorithm of
Miller supposing that our r (for example r=$3^{97}+3^{49}+1$, so
we can neglect 3 bit in front of 94 bit) is cruse.\\
Firstly, we have $t_{r}=(f_{r}(D_{Q}))^{\frac{q^{k}-1}{r}}$ =
$(\frac{f_{r,P}(Q+S)}{f_{r,P}(S)})^{\frac{q^{k}-1}{r}}$ with
$D_{Q}$ = [Q+S]-[S] for an arbitrary chosen S in the elliptic
curve concerned.
The algorithm of Miller is resumed in table 4\\
\begin{table}[h!!]
\hspace*{-3cm} \caption{first tranche}
\begin{tabular}{|c|}
 \hline
\textbf{Compute of $\frac{f_{r,P}(Q+S)}{f_{r,P}(S)}$} : first tranche\\
\hline
\small
 \hspace{+1.2cm}\textbf{Input: r =
($r_{n}$...$r_{0}$)(binary representation )},
\\ \hspace{.75cm} P $\in$ $E[r]$($\subset$ E($F_{q}$)) and
Q $\in$ $G_{1}$($\subset$ $E(F_{q^{k}}$)) \\
\hspace{-3.2cm} S $\in$ $G_{1}$($\subset$ $E(F_{q^{k}}$))\\
\small
 \hspace{-2.1cm}\textbf{Output:} $f_{r,P}(Q)$ $\in$
$G_{3}$ ($\subset$ $F_{q^{k}}^{*})$
\\ \hspace{-5.1cm} T $\leftarrow$ P\\
 \hspace{-5.1cm}  $f_{1}$ $\leftarrow$ 1 \\
 \hspace{-3.5cm} for i = n - 1 to 0 do \\
 \hspace{-4.5cm} \textbf{1:} T $\leftarrow $  [2]T \\
 \hspace{-1.84cm}  $f_{1}$ $\leftarrow$
${f_{1}}^{2}$ $\times$ $\frac{l_{1}(Q+S)}{l_{1}(S)}$ $\times$ $ \frac{v_{1}(S)}{v_{1}(Q+S)}$ \\
\hspace{-1.2cm} $l_{1}$ is the tangent to the curve in T.\\
\hspace{-.8cm} $v_{1}$ is the vertical to the curve in [2]T.\\
\hline
\end{tabular}
\end{table}\\
In this algorithm, we need three stages: (1) computation of ECDBL
(we neglect ECADD)  (2) computation of $l_{1}(Q + S), l_{1}(S),
v_{1}(Q + S), v_{1}(S)$ (3) update of $f_{1}$\\
According to [29] we have so:
 C(Miller )= r $log2(4Mu_{k} + 2Sq_{k} + (6k +
7)Mu + 7Sq)$ with r log2 is the number of iterations. If r is in
the same level of security as n, we will have: \\C(Miller )= n
log2$(4Mu_{k} + 2Sq_{k} + (6k + 7)Mu + 7Sq)$.\\\\
NB: \begin{enumerate}
    \item Even if we are basing in a work[29] made in 2003, but this
complexity is nearest to the one[30] done in 2009 section II.2.1.
And in this latter the author don't take into account $l_{1}(Q +
S)$, $v_{1}(Q + S)$, multiplication: $l_{1}(Q+S)\times v_{1}(S)$,
$l_{1}(S)\times v_{1}(Q+S)$
    \item k designs the embedding degree of the field used. For example
    $F_{p^{k}}$; $Mu_{k}$: multiplication in this field; $Sq_{k}$: squaring in this
    field.
    \item Certain work use twist which eliminate the calculate of
    $v_{1}$,  this is possible for embedding degree divided by
    2, 3, 4, 6. But we don't take it
    into consideration in this work
   \item According to[31], for k=$2^{i}3^{j}$ $Mu_{k} = 3^{i}5^{j}Mu$; $Mu_{k}$ $\sim$ $Sq_{k}$
   so $Sq_{k}$ $\cong$ $3^{i}5^{j}Mu$.
\end{enumerate}
We take k=$2^{i}3^{j}$ as an experiment embedding to make our
comparison, this because of last step : step number 4. And the fact that C(Mu)$\simeq$ C(Sq). So:\\
C(Miller )= nlog2 $((6.3^{i}5^{j}+(6k + 14)) O((log n)^{2}))$.\\
For k=12 and in a level of security =80. We have:C(Miller )=28480
Log 2O$(6400(log 2)^{2})$.\\
C(pairing)=nlog2 $((6.3^{i}5^{j}+(6k + 14)) O((log
n)^{2}))$+$(\frac{3}{2}$ lgn)$O((log n)^{2}))$\\
We move now to the inversion of two pairing:\\ According to
section 2.1.7 instead of calculate
$\frac{t_{r_{1}}(D_{r_{1}}(D_{Q_{1}}))}{t_{r_{2}}(D_{r_{2}}(D_{Q_{2}}))}$=$\frac{(f_{r_{1},P_{1}}(D_{Q_{1}}))^{\frac{q^{k}-1}{r_{1}}}}
{(f_{r_{2},P_{2}}(D_{Q_{2}}))^{\frac{q^{k}-1}{r_{1}}}}$, if
$P_{1}$ and $P_{2}$ have the same order r=$r_{1}=r_{2}$, we
calculate only $t_{r}(D_{r}(D_{Q_{1}}))$ $\times$
$t_{r}(D_{r}(D_{Q_{2}}))$. This reduce the complexity from
4$Mu_{k}$ to only 2$Mu_{k}$ (as inversion in $F_{p^{k}}$ is
approximated  to 4$Mu_{k}$ following [29])\\
Using this,  the technique proposed in the section 2.1.7 and
complexity given in [29] (first tranche), we have:\\ C(Inversion
of Tate Pairing)=nLog2(2(4Mu + 6Sq) + 2(3Mu + 1Sq) + 4(3kMu) +
$4Mu_{k} + 2Sq_{k} $) + 1C(exponent)= (28+12k + 6.$3^{i}5^{j}$)Mu+
$\frac{3}{2}$ $log n
O({logn}^{2})$=nLog2(28+12k+6.$3^{i}5^{j}$)$O((log
n)^{2})$ + $\frac{3}{2}$ $log n O({logn}^{2})$\\
We will use all this complexity in the following section when we
have ambiguity. \vspace*{-.5cm}
\subsubsection*{Efficient Classification} To classify our
cryptosystems we compared them following  each taps: Params,
Extract, Encrypt, Decypt. So we have following the complexity in
table 3 and the complexity declared in the previous section:
\\It is clear from table 3 that:$({BF-Gentry})_{Params}$ $<$
$({SK-ChenCheng})_{Params}$ \& $Water_{Params}$  $<$
$BB2_{Params}$.  To compare $BB1_{Params}$ and $Gentry_{Params}$
we will compare only 2C($Mul_{sca}$)+$C(Exp_{ffi})$ and 3C(pair).
As we have $\frac{106}{3}(n-1)$+$\frac{3}{2}$logn $<$
$(Log2^{n})$(18.$3^{i}.5^{j}$+ 3(6k+14)+$\frac{9}{2}$Logn),
$BB1_{Params}$ $<$ $Gentry_{Params}$.\\ So:
$({BF-Gentry})_{Params}$ $<$ $({SK-ChenCheng})_{Params}$ \&
$Water_{Params}$  $<$ $BB2_{Params}$ $<$ $BB1_{Params}$ $<$
$Gentry_{Params}$.
\\
For the Extract, the fact that $Mul_{sca}$ has in its formulate an
Mul and Sq multiplied by n, will help us in a more statement. The
only ambiguity that we can have is between BF and BB1, but as we
have
C(square root)$<$C(Mul) we will have:\\
$({SK-ChenCheng})_{Extract}$ $<$ $BB2_{Extact}$ $<$
$({BF-Galnido})_{Extract}$ $<$
$BB1_{Extract}$$<$ $Gentry_{Extract}$ $<$  $Water_{Extract}$.\\
In the level Encrypt we have regrouped the complexity for each
cryptosystem, using the fact that an inversion in $F_{p^{k}}$ is approximated to $4Mu_{k}$ [29] (for Gentry) we find that:\\
 $({SK-ChenCheng})_{Encrypt}$
$<$$BB1_{Encrypt}$ \& $Water_{Encrypt}$ $<$$BB2_{Encrypt}$$<$
$Gentry_{Encrypt}$ $<$
$({BF-Galnido})_{Encrypt}$. \\
As far as for the Decrypt we have:\\
$({BF-Galnido})_{Decrypt}$ $<$ $({SK-ChenCheng})_{Decrypt}$
 $<$   $Water_{Decrypt}$ $<$ $BB2_{Decrypt}$  $<$ $BB1_{Decrypt}$
$<$ $Gentry_{Decrypt}$. The classification  between
$({BF-Galnido})_{Decrypt}$ - $({SK-ChenCheng})_{Decrypt}$; as well
as $BB2_{Decrypt}$ - $BB1_{Decrypt}$ and $BB1_{Decrypt}$ -
$Gentry_{Decrypt}$ are clair. We have an ambiguity between
$Water_{Decrypt}$ and $BB2_{Decrypt}$, $Water_{Decrypt}$ and
$({SK-ChenCheng})_{Decrypt}$. But as we have
nlog2(28+12k+6.$3^{i}.5^{j}$)+1 $>$ nlog2(6.$3^{i}.5^{j}$+6k+14)+
$\frac{53}{2}$(n-1), because (14+6k)log2+1$>$ $\frac{53}{2}$(n-1)
(we can take the minimal case k=2) we can so conclude.\\
\begin{table}[h!]
\begin{center}
\caption{Classification }\vspace*{+.3cm}
\begin{tabular}{|c|c||c|c|c|c|c|}
\hline
   & $BF_{Gal}$ & $SK_{Ch-Chg}$ & BB1 & BB2 & Water & Gentry \\
  \hline
  Params & 1 & 2 & 4 & 3 & 2 & 5 \\
  \hline
  Extract & 3 & 1 & 4 & 2 & 6 & 5 \\
  \hline
  Encrypt & 5 & 1 & 2 & 3 & 2 & 4 \\
  \hline
  Decrypt & 1 & 2 & 5 & 4 & 3 & 6 \\
  \hline
  \hline
    Sum  & 10 & 6 & 15  & 12  & 13 & 20 \\
\hline
  Class & $(2^{sd})$& $(1^{st})$& $(5^{th})$ &$(3^{th})$ & $(4^{th})$ &$(6^{th})$ \\
\hline
\end{tabular}
\end{center}
\end{table}
\subsection{Final Classification}
As a consequent of all what we have seen before, we regrouped our
results in the following table:
\begin{table}[h!]
\begin{tabular}{|c|c|c|c|c|c|c|}
  \hline
   & BF & SK & BB1 & BB2 & Water & Gentry \\
\hline
  Class TABLE 1 & $(2^{sd})$ & $(5^{th})$ & $(4^{th})$ & $(6^{th})$ & $(1^{st})$ & $(3^{th})$ \\

  \hline
   Class TABLE 5  & $(2^{sd})$& $(1^{st})$& $(5^{th})$ &$(3^{th})$ & $(4^{th})$ &$(6^{th})$ \\
  \hline
 Sum &4 & 6 & 9 & 9 & 5 & 9\\
  \hline
  Final $Class_{1}$ & $(1^{st})$ & $(3^{th})$ & $(4^{th})$ & $(4^{th})$ & $(2^{sd})$ & $(4^{th})$ \\
   \hline
\end{tabular}
\end{table}
\subsection{Propriety Associate} In this section as [11] we also
enriched our study with the additional properties such as:
Multi-recipient encryption, Threshold secret sharing, Hierarchical
identities. Our comparison is totaly difference from that of [11].
Because we do not mark only the property as [11] to the crypto
systems, but we test the best
crypto system which verify the property wished.\\ We make firstly the following recall with a little details:\\
\textbf{Multi-recipient encryption (1):} Is the act of encrypting
a single message to multiples users. So this priority
requires a \textbf{small Encrypt}\\
 \textbf{Threshold secret sharing
(2):} Is the fact of dividing the key Master on several
authorities, to avoid the concentration on one. And  each of them
has the advantage to calculate a corresponding private key. So
this
priority requires a \textbf{small Extract}\\
\textbf{Hierarchical Identity (3):} Is the fact of arranging
multiples identities in the hierarchy (many authorities classify
in an hierarchy) using the same Params. So  each of the super
authority generate the corresponding key to its down. This
priority requires \textbf{Extract and Encrypt smaller}. Its
ranking is calculated as (Extract + Encrypt)
\begin{table}[h!]
\begin{tabular}{|c|c|c|c|c|c|c|}
  \hline
 & BF & SK & BB1 & BB2 & Water & Gentry \\
  \hline
M-r enc (1)& 5 & 1 & 2 & 3 & 2 & 4  \\
\hline
Th s sh (2)& 3 & 1 & 4 & 2 & 6 & 5 \\
\hline
Hi id (3)& 4 & 1 & 3 & 2 & 4 & 5 \\
\hline
Sum & 12 & 3 & 9 & 7 & 12 & 14 \\
\hline
 $Class_{2}$    & $(4^{th})$ & $(1^{st})$ & $(3^{sd})$ & $(2^{th})$ & $(4^{th})$ & $(5^{th})$\\
 \hline
\hspace{-.14cm}Specific $Class_{Fi}$ = \\ $Class_{1}$ +$Class_{2}$
\hspace{-.3cm}
& $(2^{st})$ & $(1^{st})$ & $(4^{sd})$ & $(3^{th})$ & $(3^{st})$ & $(5^{th})$\\
\hline
\end{tabular}
\end{table}
\section{Second goal}
In the following sections, we will give an efficient schemes
IBE/HIBE in the model selective ID. A comparison in terms of
performance and complexity with BB1 and BBG scheme is in favor of
our scheme.
\subsection{Preliminaries}
To be familiarized with the difference between IBE and HIBE, we
give in the following the functionality of each others.
\subsubsection{Functionality of IBE:} An IBE system contains four
basic components in its
construction:\\
\textbf{Setup:} A trusted central authority manages the parameters
with which keys are created. This authority is called the Private
Key Generator or PKG. The PKG takes a security parameter k and
returns \textbf{params} (system parameters) and
\textbf{master-key}. The system parameters will be publicly known,
while the master-key
will be known only to the (PKG).\\
\textbf{Extract:} Takes as input \textbf{params},
\textbf{master-key}, and an arbitrary $ID_{R}$,  it returns a
private key $d_{ID_{R}}$.\\
\textbf{Encryption:} When Alice wishes to encrypt a message to
Bob, he encrypts the message to him by computing or obtaining the
public key, and then encrypting a plaintext message M
with params, $ID_{Bob}$ to obtain ciphertext C.\\
\textbf{Decryption:} When Bob has C, he contact the PKG to obtain
the private key $S_{Bob}$, he decrypts C to obtain the plaintext
message M. \subsubsection{ IBE security notions} As it was known
Boneh and Franklin define  in [2] a chosen ciphertext security for
IBE systems under a chosen identity attack. In this model the
adversary is allowed to adaptively chose the public key it wishes
to attack. In [13] Canetti, Halevi, and Katz  define another
notion it is a weaker notion of security. In this model the
adversary commits ahead of time to the public key
it will attack.\\
 Before giving its functionality we recall firstly that the security of a cryptographic scheme combining the possible goals and attack models. The most important
goal are: indistinguishability (IND/sIND), Semantic Security.
Regarding attacks we have: chosen-plaintext attacks (CPA),
chosen-ciphertext
attacks (CCA). The relation between all this was given in [32].\\
\textbf{Definition:IND-{ID/sID}-\{CCA, CPA\}} \begin{description}
    \item[]\hspace*{.5cm}  Let $\Gamma$
= (S,X,E,D) be an IBE scheme, and let A = $(A_{0},A_{1},A_{2})$ be
any 3-tuple of PPT oracle algorithms. For ATK = ID/sID-CPA,
ID/sID-CCA, we say $\Gamma$ is IND/sID-ATK secure if for any
3-tuple of PPT oracle
algorithms A,$|$ $\wp$r(1)-$\wp$r(2) $|$ $\in neg$ , where\\
$\wp$r(i)=
$\left \{%
\begin{array}{ll}
    v=0
     \left |%
     \begin{array}{ll}
    \hbox{(id,$\gamma)\longleftarrow A_{0}(1^{l})$} \\
 \hbox{(pms,mk) $\longleftarrow S(1^{l})$;}\\
  \hbox{$((m^{(1)},m^{(2)}, id_{ch}), \sigma) \longleftarrow A_{1}^{O_{1},O_{2}}
(pms, id, \gamma)$}\\
\hbox{$c \longleftarrow E(pms, id_{ch},m^{(i)}); $}\\
\hbox{$v \longleftarrow A_{2}^{O_{1},O_{2}}(\sigma, (id_{ch}, c))$}\\
\end{array}%
\right.
\end{array}%
\right \} $.\\
The expression represent the oracles $O_{1},O_{2}$. Additionally,
$m^{(1)}$ and $m^{(2)}$ are required to have the same length;
neither $A_{1}$ nor $A_{2}$ are allowed to query $O_{1}$ on the
challenge identity $id_{ch}$, and $A_{2}$ can not query $O_{2}$ on
the challenge pair ($id_{ch}$, c). These queries may be asked
adaptively (like CCA2 after phase 2), that is, each query may
depend on the answers obtained to the previous queries.
\end{description}
\subsubsection{ Functionality of HIBE}
 Like IBE system, the
Hierarchical Identity Based Encryption (HIBE) system consists of
four algorithms [15][16]: \textbf{Setup}, \textbf{KeyGen},
\textbf{Encrypt}, \textbf{Decrypt}.\\ In HIBE, however, identities
are vectors, a vector of dimension k represents an identity at
depth k. The Setup algorithm generates system parameters, denoted
by params, and a master key master-key. We refer to the master-key
as the private key at depth 0 and note that an IBE system is a
HIBE where all identities are at depth 1. Algorithm KeyGen takes
as input an identity ID = $(I_{1}$, . . . , $I_{k})$ at depth k
and the private key $d_{ID}|k-1$ of the parent identity ID$|$k-1 =
$(I_{1}$, . . . , $I_{k-1})$ at depth k -1, and then outputs the
private key $d_{ID}$ for identity ID. The encryption algorithm
encrypts messages for an identity using params and the decryption
algorithm decrypts ciphertexts using the private key.
\subsubsection{The main approach of IBE}
We can classify the cryptosystems of IBE in three categories:
 \begin{description}
    \item[$\bullet$] Full-Domain-Hash approach: In this
model we project in the elliptic curve instead of the finite
field, its prototype is summarized by the idea of Boneh-Franklin
[2].
\end{description}
\begin{description}
    \item[$\bullet$ $\bullet$] Exponent-Inversion approach: In this approach
the identity key to be used in the Extract is as an inverse. The
second scheme of Boneh-Boyen (BB2)[4],  that's of Sakai-Kasahara
(SK) [3], also Gentry [6] work with this approach.
\end{description}\begin{description}
    \item[$\bullet$ $\bullet$ $\bullet$] Commutative-Blinding approach, defined by the first IBE scheme of Boneh-Boyen
(BB1)[4]. It is based on the idea of creating, from two or more
secret coefficients, two blinding factors that "commute" with each
other under the pairing. The main quality that characterize this
paradigm is the greater flexibility provided by its algebraic
structure. Since the identity presented in the Extract is in the
form linear.
\end{description}
\subsubsection{ Selective Identity  IBE/HIBE  Security Notions}
Selective Identity  for an IBE function as follow, but we give
only version CPA i.e without using the
extraction decrypt queries in phase 1:\\
\textbf{Init:}
\begin{description}
    \item[] \hspace*{.5cm} The adversary outputs an identity $ID^{*}$ where it wishes to be
challenged.
\end{description} \textbf{Setup:}
\begin{description}
    \item[] \hspace*{.5cm} The challenger runs the
Setup algorithm. It gives the adversary the resulting system
parameters params. It keeps the master-key to itself.
\end{description}
\textbf{Phase 1:}
\begin{description}
    \item[] \hspace*{.5cm} The adversary issues queries $q_{1}, . . . , q_{m}$ where query
$q_{i}$ is:
\begin{itemize}
    \item
 Private key query $ <ID_{i}>$ where
$ID_{i} \neq ID^{*}$ and $ID_{i}$ is not a prefix of $ID^{*}$. The
challenger responds by running algorithm KeyGen to generate the
private key $d_{i}$ corresponding to the public key $ <ID_{i}>$.
It sends $d_{i}$ to the adversary.
\end{itemize}
\end{description}
\textbf{Challenge:}
\begin{description}
    \item[]
\hspace*{.5cm} Once the adversary decides that Phase 1 is over it
outputs two equal length plaintexts $M_{0},M_{1} \in \verb"M"$ on
which it wishes to be challenged. The challenger picks a random
bit b $\in$ \{0, 1\} and sets the challenge ciphertext to C =
Encrypt(params, $ID^{*},M_{b}$). It sends C as the challenge to
the adversary.\end{description} \textbf{Phase 2:}
\begin{description}
    \item[] \hspace*{.5cm} As phase 1
\end{description}
\textbf{Guess:}
\begin{description}
    \item[] \hspace*{.5cm} Finally, the adversary outputs a guess $b_{0} \in$ \{0, 1\}.
The adversary wins if $b = b_{0}$.
\end{description}
 We refer to such an adversary A as an IND-sID-CPA
adversary. We define the advantage of the adversary A in attacking
the scheme E as $Adv_{\varepsilon,A}$ = $|$ Pr[b = $b_{0}$] -
$\frac{1 }{2}$ $|$ \ The probability is over the random bits used
by the challenger and the adversary.\\
We say that an IBE (or HIBE ID = $ID_{1}, ID_{2}, ..., ID_{k}$ for
a level k) system E is (t, $q_{ID},
\varepsilon)$-selective-identity, adaptive plaintext secure if for
any IND-sID-CPA adversary A that runs in time t, makes at most
$q_{ID}$ chosen private-key queries, we have that
$Adv_{\varepsilon,A}$ = $|$ Pr[b = $b_{0}$] - $\frac{1 }{2}$ $|$
$<$ $\varepsilon$. \subsubsection{ Selectiv$e^{+}$-ID Model} In
Selectiv$e^{+}$-ID [14] we give a more power
 to the adversary. The power is a modification that will be given in the Challenge
 phase (prefix of the $ID^{*}$).\\
\textbf{Challenge:} A outputs two equal length messages
$M_{0},M_{1}$ and an identity v+ where v+ is either $ID^{*}$ or
any of its prefixes. In response it receives an encryption of M
 under v+, where
 is chosen uniformly
at random from \{0, 1\}. This model is more general than the sID
model, because  the adversary is allowed to ask for a challenge
ciphertext not only on $ID^{*}$ but also on any of its prefixes.\\
A protocol secure in the selectiv$e^{+}$-ID model is obviously
secure in the selective-ID model.
\subsubsection{Problem Bilinear of Diffie Hellman Assumption}
During all the following section, we use the multiplicative
expression instead of the additive one to simplify the proof of
security. So we will give the following definition in  the
multiplicative
expression.\\
\textbf{Definition 8}:\begin{description}
    \item[] \hspace*{.5cm}((Decisional) Bilinear Diffie-Hellman Problem DBDHP). Let $G_{1}$, $G_{2}$ two rings with prime order  q. Let
\\\^e : $G_{1} \times G_{2}$ $\longrightarrow$ $G_{T}$ be an
application admissible and bilinear and let g be a generator of
$G_{1}$. The DBDHP in $<$ $G_{1},G_{2}$, \^e $>$ is so: Given $<$
g, $g^{a}, g^{b}, g^{c}$, z $>$ for a, b, c $\in$ $Z_{q}$ and z
$\in G_{2}$. we say that an algorithm \textsl{A} that outputs b
$\in$ \{0,1\} has advantage $\varepsilon$ in solving the decision
BDHP in G if:\\
    $|$ Pr [ g, $g^{a}, g^{b}, g^{c}$, \^e($g,g)^{abc}$ ]-Pr [g, $g^{a}, g^{b}, g^{c}$, z ]$|$ $>$ $ \varepsilon $ \\
where the probability is over the random choice of generator g in
$G_{1}$, the random choice of a, b, c in $Z_{q}$, the random
choice of z $\in G_{2}$, and the random bits of \textsl{A}. The
distribution on the left is refereed as $\textsl{P}_{BDHP}$ and
the distribution on the right as $\textsl{R}_{BDHP}$.
\end{description}
\textbf{Definition 9:}\begin{description}
    \item[] \hspace*{.5cm}((Decisional)k-Bilinear Diffie Hellman Inversion Problem (Dk-BDHIP)).
Let
 k be an integer, and x $\in$ $Z_{q}^{*}$, $ g \in G_{2}^{*}$, \^e : $G_{1} \times G_{2}$ $\longrightarrow$ $G_{T}$, $T \in
 G_{T}$.
 Can we make the following separation:\\
$|$ Pr [ g, $g^{x}, g^{x^{2}},..., g^{x^{k}}$,
\^e($g,g)^{\frac{1}{x}}$ ]-
    Pr [g, $g^{x}, g^{x^{2}},..., g^{x^{k}}$, T ]$|$ $>$ $\varepsilon$  \\
\end{description}
\textbf{Definition 10:}\begin{description}
    \item[] \hspace*{.5cm}((Decisional)k-Weak Bilinear  Diffie Hellman Inversion Problem ($Dk-wBDHIP^{*}$)).
Let
 k be an integer, and x $\in$ $Z_{q}^{*}$, $ g \in G_{2}^{*}$, \^e : $G_{1} \times G_{2}$ $\longrightarrow$ $G_{T}$, $T \in
 G_{T}$.
 Can we make the following separation:\\
$|$Pr [ g, h, $g^{x}, g^{x^{2}},..., g^{x^{k}}$,
\^e($g,h)^{x^{\frac{1}{x}}}$ ]-
    Pr [g, h, $g^{x}, g^{x^{2}},..., g^{x^{k}}$, T ]$|$ $>$  $ \varepsilon $

\end{description}
\subsection{Efficient IBE} Our second goal behind this work is to
represent  an efficient scheme in the model selective ID. This
notion of security is weaker, Boneh et al prove that to pass from
selective ID to full domain we will introduce a factor N.
Additionally, as we have seen previously the BB1 is also more
complex. We propose so to reduce this scheme or rather to propose
a scheme in the approach Commutative Blinding and under the model
Selective ID more reduced.
\subsubsection{ Construction} To avoid the use of two pairing in the Decrypt as with BB1, we collect in our approach the
principal of the inverse in Extract as with BB2[4] and that's of
the commutative Blinding[10], our procedure is as
follow:\\\\\\\\\\\\\\\\\\\\\\\\\\\\\\\\\\
\begin{table}[h!]
\begin{tabular}{|c|}
  \hline
  \normalsize Our Scheme \\
  \hline
 \hspace*{-1.2cm}\textbf{Setup}. Let ($G_{1}, G_{T})$
a bilinear group. Choose a generator $g\in
G_{1}$ \\ \hspace*{2.cm}  and set $P_{pub_{1}} = g^{l}$ $\in {G_{1}}^{\star}$. Calculate $e(g,g)=x$ and $e(g,g)^{a}=x^{a}=y$.\\
\hspace*{+1.4cm} $M_{pk}$= \{$G_{1},  G_{T},$ $P_{{pub}_{1}}$, x,
y \}.  The Master secret key is $M_{sk}$= \{l,a\}
\\
 \hspace*{+1.6cm} Message space is  $\{0, 1\}^{n}$, ciphertext space is ${G_{1}}^{*}$
$\times \{0, 1\}^{n} \times \{0, 1\}^{n}$.
\\\hspace*{-.9cm} \textbf{ Extract:} Given an identifer  $ID_{A}$ $\in$
$\{0,1\}^{n}$  of entity A, $M_{pk}$  and $M_{sk}$\\
\hspace*{+1.9cm}  Pick an $r_{ID_{A}}$ $\in Z_{q}$,  returns
$g^{\frac{a+ID_{A}}{r_{ID_{A}}l}}$=$g^{\frac{\frac{a}{r_{ID_{A}}}+r'_{ID_{A}}ID_{A}}{l}}$=$g^{\frac{a'+r'_{ID_{A}}ID_{A}}{l}}$,
$d_{A}$= $(r_{ID_{A}},g^{\frac{a+ID_{A}}{r_{ID_{A}}l}})$\\
\textbf{ Encrypt:} Given a m $\in$ M , $ID_{A}$ and $M_{pk}$, the
following step are
formed:\\\hspace*{-4.6cm} 1. Pick a random s in $Z_{q}$\\\
\hspace*{-.7cm} 2.Compute
$z^{s(ID_{A}+a)}$=$e(g,g)^{s(ID_{A}+a)}=(x^{ID_{A}}y)^{s}$ \\
\hspace*{+1.1cm}Set the ciphertext to be
$C=(g^{ls}={P_{pub_{1}}}^{s},m.z^{s(ID_{A}+a)})$\\
\hspace*{+.8cm} \textbf{Decrypt:} Given a ciphertext C =
(u,v)$\in$$\textsl{C}$, $ID_{A}$, $d_{A}$ and $M_{pk}$, follow the steps\\
\hspace*{-1.4cm} 1. Compute e$(u^{r},d_{A})$ and  output m=$\frac{v}{e(u^{r_{ID_{A}}},g^{\frac{a+ID_{A}}{r_{ID_{A}}l}})}$\\
  \hline
\end{tabular}
\end{table}\\
\normalsize Firstly it is necessary to a fix a security parameter
t. l and a follow the degree of security of this parameter.
\vspace*{-.4cm}
\subsubsection*{\textbf{Correctness}} As we have:\\
e$(u^{r_{ID_{A}}},g^{\frac{a+ID_{A}}{r_{ID_{A}}l}}))=e(g^{lsr_{ID_{A}}},g^{\frac{a+ID_{A}}{r_{ID_{A}}l}})$=$e(g,g)^{s(ID_{A}+a)}$,
our scheme is then correct \vspace*{-.4cm}
\subsubsection*{Observation} In our scheme we use the
master key (s,a,$\hat{P}$=$\frac{1}{s}P_{2}$),  the private key
will be $d_{A}$=$(r_{ID_{A}}(a+H_{1}(ID_{A})))\hat{P}$. As a
consequence
 the
$\hat{P}$ in our scheme will be computed one time and will be
reuse to each demands, contrary to [4]. Noting that the syntax
$d_{A_{1}}$ of a given entity $A_{1}$,  we couldn't calculate the
private key $d_{A_{2}}$ for another entity $A_{2}$, because we
don't know a and we cannot inverse s. Also we change $r_{ID_{A}}$
for each Identity. \vspace*{-.45cm}
\subsubsection{ Prove of Security} Before proving the security of
our scheme, we note that $k^{-}$-BDHI, mean that we can use any
\\k $>$ 0 (it is not linked to the number of users as with[4]).
And it is of our choice (we can choose it 2 or any number), by
contrast with [4]
we need at lest $2^{50}$ (after [7]) for a 80 level of security.\\
The security of our scheme is basing on $Dk^{-}$-BDHI assumption
since:\\ \textbf{Theorem:} Suppose the (t, $k^{-},
\varepsilon$)-Decision BDHI assumption holds in G of size $|$G$|$
= p. Then our scheme is $(t', q_{S}, \varepsilon)$-selective
identity, chosen plaintext (IND-sID-CPA) secure, with an
advantage:\\ \textsl{adv}$^{scheme}$(t') $>$
\textsl{adv}$^{Dk^{-}-DBDHIP}$(t-O($\tau$ q)) for any $q_{S}$ $
<$ q . Where $\tau$ is the time needed for an exponentiation in the following study.\\
\textbf{Proof.} Suppose A has advantage $\varepsilon$ in attacking
our scheme. We build an algorithm B that uses A to solve the
Decision $k^{-}$-BDHI problem in G. Algorithm B is given as input
a random ($k^{-}$+2)-tuple \\$(g, g^{\alpha}, g^{\alpha^{2}},..
g^{\alpha^{k^{-}}}, T) \in$ $G_{1}^{k^{-}+1}\times G_{T}$ that is
either sampled from $\textsl{P}_{BDHI}$ (where $T = e(g,
g)^{1/\alpha}$) or from $\textsl{R}_{BDHI}$ (where T is uniform
and independent in $G_{T}$). The goal of the algorithm B is to
output 1 if $T = e(g, g)^{1/\alpha}$ and 0 otherwise. Algorithm B
works by interacting with A in a selective identity game as
follows:\\\\
 \textbf{Setup.}\begin{description}
    \item[] \hspace*{.5cm} To generate the system parameters, algorithm B
does the following:\\
 In the beginning algorithm A give B the identity $I^{*}$=$\frac{a_{1}}{b_{1}}$ that it intends to attack.
 The selective identity game begins, but algorithm B need to prepare to it the following
 step:\\
 \textbf{Preparation step} \\
In the preparation step algorithm B choose an arbitrary x he compute $b_{1}x$\\
After he compute (implicitly): $f(\alpha)={\sum_{i=1}}^{k^{-}}c_{i}\alpha^{i}$ \\
He choose an arbitrary $r_{0}$ then he compute (implicitly) $r_{1}=r_{0}{\sum_{i=1}}^{k^{-}}c_{i}\alpha^{i-1}$\\
In the end he compute h=$g^{f(\alpha)}$ and he publish this h
\end{description}
\textbf{Phase 1:} \begin{description}
    \item[]\hspace*{.5cm}  A issues at most $q_{S}$ private key queries,
with $q_{S}$ $<$ q. Consider the i-th query for the private key
corresponding to public key $ID_{i} \neq ID^{*}$.\\ We need to
respond with a private key (r,
$h^{\frac{a+r(I-I^{*})}{\alpha}}$)\\
The I represent a general identity ID and $I^{*}$
represent an identity to be attacked \\
r is uniformly distributed in $Z_{p}$.\\ Algorithm B responds to
the query as follows:\\
Firstly it is possible that the private key in our scheme may has
the syntax $d_{A}$=$g^{\frac{a+ID_{A}}{l}}$ instead of
$d_{A}$=$g^{\frac{a+ID_{A}}{rl}}$=$g^{\frac{a}{rl}}+\frac{r'ID_{A}}{l}$=$g^{\frac{a'+r'ID_{A}}{l}}$.
But we need this latter to simplify the proof
\\
B pose R=$\frac{x}{r_{0}}+r_{1}$ he can calculate implicitly
\begin{description}
    \item[] \hspace*{.2cm} $R=\frac{f(\alpha)}{f(\alpha)}(\frac{x}{r_{0}}+\frac{r_{1}}{I-I^{*}}I-I^{*})\\
=\frac{f(\alpha)}{\alpha
{\sum_{i=1}}^{k^{-}}c_{i}\alpha^{i-1}}(\frac{x}{r_{0}}+\frac{r_{1}}{I-I^{*}}(I-I^{*}))$\\
=$\frac{f(\alpha)}{\alpha}
(\frac{x}{r_{0}{\sum_{i=1}}^{k^{-}}c_{i}\alpha^{i-1}}+\frac{r_{1}}{{\sum_{i=1}}^{k^{-}}c_{i}\alpha^{i-1}(I-I^{*})}(I-I^{*}))$\\
=$\frac{f(\alpha)}{\alpha}
(\frac{x}{r_{0}{\sum_{i=1}}^{k^{-}}c_{i}\alpha^{i-1}}+\frac{r_{0}{\sum_{i=1}}^{k^{-}}c_{i}\alpha^{i-1}}{{\sum_{i=1}}^{k^{-}}c_{i}\alpha^{i-1}(I-I^{*})}(I-I^{*}))$\\
=$\frac{f(\alpha)}{\alpha}
(\frac{x}{r_{0}{\sum_{i=1}}^{k^{-}}c_{i}\alpha^{i-1}}+
\frac{r_{0}}{I-I^{*}}(I-I^{*}))$\\
=$\frac{f(\alpha)}{\alpha}(a'+r'(I-I^{*}))$
\end{description}
With r'=$\frac{r_{0}}{I-I^{*}}$ which is easy to calculate by B\\
But a'=$\frac{x}{r_{0}{\sum_{i=1}}^{k^{-}}c_{i}\alpha^{i-1}}$ is
not it is a Master key for B like $\alpha$.\\
 \begin{description}
    \item[] \textbf{NB:} (For the master key a, A can publish $g^{a}$ in system of
parameters. To remove this a, \\ \hspace*{+.3cm} B search for an
$\sigma$ such that: $g^{a}g^{\sigma}=g^{\alpha}$)
\end{description}

So B can calculate easily $g^{R}$ as he know $g^{\frac{x}{r_{0}}}$
and
$g^{r_{0}}$\\
But $g^{R}=g^{\frac{f(\alpha)}{\alpha}(a'+r'(I-I^{*}))
}$=$h^{\frac{a'+r'(I-I^{*})}{\alpha}} $ which is a valid private
key and so B can give A the private key (r',$h^{\frac{a'+r'(I-I^{*})}{\alpha}} $)\\
More B has not the advantage to calculate the private key for
$I^{*}$
\end{description}
\textbf{Challenge.}
\begin{description}
    \item[] \hspace*{.4cm} A outputs two messages $M_{0}, M_{1} \in G_{1}$. Algorithm B
picks a random bit b $\in$ \{0,1\} and a random l' $\in$
${Z_{p}}^{*}$. It responds with the ciphertext prepared as
follow:\\
He have  $h^{s}=h^{\frac{s}{\alpha}.\alpha}=h^{l'\alpha}$ = $c_{1}$, with l'=$\frac{s}{\alpha}$\\
And $c_{2}$=$MT_{h}^{\frac{s(xb_{1}+a_{1}}{b_{1}})}=T_{h}^{s(x+I^{*})}$ (or rather $c_{2}$=$MT_{h}^{\frac{s(ab_{1}+a_{1}}{b_{1}})}=T_{h}^{s(a+I^{*})}$)\\
So if $T_{h}=e(h,h)^{\frac{1}{\alpha}}$ he will have
$e(h,h)^{\frac{s}{\alpha}(x+I^{*})}=c_{2}=e(h,h)^{l'(x+I^{*})}$\\
And he combine
CT=$(c_{1},c_{2})=(h^{l'\alpha},e(h,h)^{l'(x+I^{*})})$ which is a
valid ciphertext under $ID^{*}$\\
If $T_{h}$ is uniform in $G_{1}$, then CT is independent of the
bit b.\\
\end{description}
 \textbf{Phase 2.} \begin{description}
    \item[] \hspace*{.4cm} A issues more private key queries, for a total
of at most $q_{S}$ $<$ q. Algorithm B responds as before.
\end{description}
\textbf{Guess.} \begin{description}
    \item[] \hspace*{.4cm} Finally, A outputs a guess b' $\in$ \{0, 1\}. If b =
    b'
then B outputs 1 meaning T = $e(g, g)^{\frac{1}{\alpha}}$.
Otherwise, it outputs 0 meaning T $\neq$ $e(g,
g)^{\frac{1}{\alpha}}$.
\end{description}
When the input $k^{-}+2$-tuple is sampled from
$\textsl{P}_{BDHIP}$ (where T = $e(g,g)^{\frac{1}{\alpha}}$) then
A's view is identical to its view in a real attack game and
therefore A must satisfy $|$Pr[b = $b'$] - 1/2$|$ $>$
$\varepsilon$. On the other hand, when the input $k^{-}+2$-tuple
is sampled from $\textsl{R}_{BDHIP}$ (where T is uniform in
$G_{T}$) then Pr[b = $b'$] = 1/2. Therefore, with g uniform in
$G_{1}$, T uniform in $G_{T}$ we have that:\\
$ \left |%
     \begin{array}{ll}
    \hbox{Pr [ g, $g^{\alpha}, g^{\alpha^{2}},..., g^{\alpha^{k^{-}}}$, \^e($g,g)^{\frac{1}{\alpha}}$ ]-
    Pr [g, $g^{\alpha}, g^{\alpha^{2}},..., g^{\alpha^{k^{-}}}$, T ] } \\
\end{array}%
\right |$ $\geq$
$ \left |%
     \begin{array}{ll}
    \hbox{($\frac{1}{2}\pm \varepsilon)-\frac{1}{2}$=$\varepsilon$}
\end{array}%
\right |.$ \hspace*{2cm} $\Box$
\\
Noting that in IBE, $s^{+}$-ID and s-ID are the same, the
difference may be in HIBE. \vspace*{-.35cm}
\subsubsection{ Discussion}
\subsubsection*{ $\blacktriangleright$ Comparison with BB1 and BB2}
In the following we compare the efficiency of our scheme with BB1
 (version IBE[11] but under selective ID) and with BB2. We have seen
 above that we make a little change in BB2. This change is effective as we reduce
the complexity of BB2. More than that our scheme is also more
efficient than BB1(version IBE[11]). All this statements are
summarized in  table 6. \vspace*{-.3cm}
\subsubsection*{$\blacksquare $ Compute of complexity}
\begin{table}[h!]
\caption{}
\begin{tabular}{|c|c|}
  \hline
   & BB1  \\
  \hline
  Params & \footnotesize 2$Exp_{ffi_{{G_{1}/Z_{q}}}}$+1Pair+$1Exp_{ffi_{G_{T}/Z_{q}}}$  \\
  \hline
  Extract & \footnotesize $2Mul_{ffi_{{Z_{q}/Z_{q}}}}+2Exp_{ffi_{{G_{1}/Z_{q}}}}$  \\
  \hline
  Encrypt & \footnotesize $1Mul_{ffi_{{Z_{q}/Z_{q}}}}$+$3Exp_{_{{G_{1}/Z_{q}}}}$+$1Exp_{ffi_{{G_{T}/Z_{q}}}}$\\
  \hline
  Decrypt & \footnotesize 2Pair+$1Div_{ffi_{{G_{T}/G_{T}}}}$  \\
  \hline
  Sum & \footnotesize 3Pair+$1Div_{ffi_{{G_{T}/G_{T}}}}+3Mul_{ffi_{{G_{1}/G_{1}}}}+7Exp_{ffi_{{G_{1}/Z_{q}}}}+2Exp_{ffi_{{G_{T}/Z_{q}}}}$ \\
  \hline
\end{tabular}\\
\begin{tabular}{|c|c|}
  \hline
   & BB2  \\
  \hline
  Params & \footnotesize 2$Exp_{ffi_{{G_{1}/Z_{q}}}}$+1Pair \\
  \hline
  Extract & \footnotesize $1Mul_{ffi_{{Z_{q}/Z_{q}}}}+1Inv_{ffi_{{Z_{q}/Z_{q}}}}+1Exp_{ffi_{{G_{1}/Z_{q}}}}$  \\
  \hline
  Encrypt & \footnotesize $1Mul_{ffi_{{Z_{q}/Z_{q}}}}$+$3Exp_{ffi_{{G_{1}/Z_{q}}}}$+$1Exp_{ffi_{{G_{T}/Z_{q}}}}$+$1Mul_{ffi_{{G_{1}/G_{1}}}}$\\
  \hline
  Decrypt & \footnotesize  1Pair+$1Div_{ffi_{{G_{T}/G_{T}}}}+1Mul_{ffi_{{G_{1}/G_{1}}}}+1Exp_{ffi_{{G_{1}/Z_{q}}}}$ \\
  \hline
  \hline
  Sum & \footnotesize 2Pair+$1Div_{ffi_{{G_{T}/G_{T}}}}+2Mul_{ffi_{{G_{1}/G_{1}}}}+7Exp_{ffi_{{G_{1}/Z_{q}}}}$+$1Inv_{ffi_{{Z_{q}/Z_{q}}}}+2Mul_{ffi_{{G_{1}/Z_{q}}}}$ \\
  \hline
\end{tabular}\\
\begin{tabular}{|c|c|}
  \hline
    & Our \\
  \hline
  Params  & \footnotesize 1$Exp_{ffi_{{G_{1}/Z_{q}}}}$+1Pair+$1Exp_{ffiG_{T}/Z_{q}}$ \\
  \hline
  Extract  & \footnotesize $1Exp_{ffi_{{G_{1}/Z_{q}}}}$+$2Mul_{ffi_{{Z_{q}/Z_{q}}}}$+$1Inv_{ffi_{{Z_{q}/Z_{q}}}}$ \\
  \hline
  Encrypt  & \footnotesize
  $1Mul_{ffi_{G_{T}/G_{T}}}$+$2Exp_{ffi_{{G_{T}/Z_{q}}}}$+$1Exp_{_{{G_{1}/Z_{q}}}}$\\
  \hline
  Decrypt  & \footnotesize 1Pair+$1Div_{ffi_{{G_{T}/G_{T}}}}$+$1Exp_{ffi_{{G_{1}/Z_{q}}}}$ \\
  \hline
  Sum & \footnotesize 2Pair+$1Div_{ffi_{{G_{T}/G_{T}}}}+2Mul_{ffi_{{Z_{q}/Z_{q}}}}+1Mul_{ffi_{{G_{T}/G_{T}}}}$+
  $3Exp_{ffi_{{G_{1}/Z_{q}}}}$+$3Exp_{ffi_{G_{T}/Z_{q}}}$+$1Inv_{ffi_{{Z_{q}/Z_{q}}}}$ \\
  \hline
\end{tabular}
\end{table}
With the fact that:\\ For example $Exp_{ffi_{* / **}}$:
Exponentiation in the finite field involved in */**,  the * is the
base of exponentiation, until the ** base of the exponent; Pair:
Pairing; Inv: Inverse; Mul: Multiplication.\\
As we have: $Complexity_{BB1} - Complexity_{Our}$ =\\
(3Pair+$1Div_{ffi_{{G_{T}/G_{T}}}}+3Mul_{ffi_{{G_{1}/G_{1}}}}+7Exp_{ffi_{{G_{1}/Z_{q}}}}+2Exp_{ffi_{{G_{T}/Z_{q}}}}$)
-\\(2Pair+$1Div_{ffi_{{G_{T}/G_{T}}}}+2Mul_{ffi_{{Z_{q}/Z_{q}}}}+1Mul_{ffi_{{G_{T}/G_{T}}}}+
3Exp_{ffi_{{G_{1}/Z_{q}}}}$+$3Exp_{ffi_{G_{T}/Z_{q}}}$+$1Inv_{ffi_{{Z_{q}/Z_{q}}}}$)
=\\1Pair+4$Exp_{ffi_{{G_{1}/Z_{q}}}}$+$3Mul_{ffi_{{G_{1}/G_{1}}}}$-$1Inv_{ffi_{{Z_{q}/Z_{q}}}}-
2Mul_{ffi_{{Z_{q}/Z_{q}}}}-1Mul_{ffi_{{G_{T}/G_{T}}}}$-$1Exp_{ffiG_{T}/Z_{q}}$
$>>$ 0\\
And we have:
$Complexity_{BB2} - Complexity_{Our}$ =\\
(2Pair+$1Div_{ffi_{{G_{T}/G_{T}}}}+2Mul_{ffi_{{G_{1}/G_{1}}}}+1Exp_{ffiG_{T}/Z_{q}}$+7$Exp_{ffi_{{G_{1}/Z_{q}}}}
+1Inv_{ffi_{{Z_{q}/Z_{q}}}}+2Mul_{ffi_{{G_{1}/Z_{q}}}}$)-\\(2Pair+$1Div_{ffi_{{G_{T}/G_{T}}}}+2Mul_{ffi_{{Z_{q}/Z_{q}}}}+1Mul_{ffi_{{G_{T}/G_{T}}}}+
3Exp_{ffi_{{G_{1}/Z_{q}}}}$+$3Exp_{ffi_{G_{T}/Z_{q}}}$+$1Inv_{ffi_{{Z_{q}/Z_{q}}}}$)
 =\\
4$Exp_{ffi_{{G_{1}/Z_{q}}}}+1Mul_{ffi_{{G_{1}/G_{1}}}}+2Mul_{ffi_{{G_{1}/Z_{q}}}}-2Mul_{ffi_{{Z_{q}/Z_{q}}}}$-$2Exp_{ffi_{G_{T}/Z_{q}}}$
$>>$ 0\\
Our scheme is then efficient than BB1 and BB2. Noting that in our
scheme and BB2, we have taking into consideration the use of r
which we need it only in the proof. The $\frac{1}{s}$ is calculate
one time and we ruse its calculate for each demand.
\vspace*{-.45cm}
\subsubsection*{ $\blacksquare $ Concrete Comparison: Technique of Boyen}
Using the technique (or rather the base) of Boyen [11], we obtain
so the following result. But, to balance the comparison between
the scheme, we consider that BB1 functions with symmetric pairing
as our scheme and BB2.\\
\small

SS @ 80-bit security level\\
  \begin{tabular}{|c|c|c|c|}
  \hline
      & BB1   \hspace*{.15cm} & BB2   \hspace*{.25cm}& Our \\
\hline \hline
Extract:     & 4   \hspace*{.15cm} & 2   \hspace*{.25cm}&  2\\
\hline \hline
Encrypt:   & 108   & 108  &  106 \\
\hline \hline
Decrypt:   & 320   & 222  & 222\\
  \hline \hline
  Sum & 432   & 332  & 330 \\
  \hline
\end{tabular}\\\\
MNT @ 80-bit security level\\
  \begin{tabular}{|c|c|c|c|}
  \hline
      & BB1   \hspace*{.15cm} & BB2   \hspace*{.25cm}& Our \\
\hline \hline
Extract:     & 0,4   \hspace*{.15cm} & 0,2   \hspace*{.25cm}&  0,2\\
\hline \hline
Encrypt:   & 100 ,8   & 100,8 & 100,6 \\
\hline \hline
Decrypt:   & 320   & 220,2  & 220,2 \\
  \hline \hline
  Sum & 421,2   & 321,2  & 321 \\
  \hline
\end{tabular}\\
\normalsize
SS: Curve Supersingular\\
MNT: Curve MNT\\
So according to these result, our scheme is more efficient than
BB1. It's complexity is nearest to BB2, but we will confirms that
our scheme is efficient than BB2. As this latter is basing in its
study of simulation in Dk-BDHIP, with k is linked to the request
identity. By contrast, our scheme is basing in D$k^{-}$-BDHIP,
$k^{-}$ $<<$ k. So our scheme is more efficient than BB2 according
to the result od Cheon[20]. \vspace*{-.3cm}
\subsection{Efficient HIBE}
\vspace*{-.2cm}
\subsubsection{Our Construction}
As we have cited above, Boneh ,Boyen and Goh [17] have proposed an
efficient scheme. This scheme reduce the ciphertext of an HIBE
from k parameters to a shorten one of only three parameters. And
the Decrypt from k product of pairing, to only two pairing. But
[17] necessitate that the use of the identity to be chosen will be
taken in $Z_{q}^{*}$ which limit the selection of the identity,
more than that [17] doesn't support the selective$^{+}$ID. In the
following proposition we overcome all this weakness.\\
\begin{table}[h!]
\begin{tabular}{|c|}
  \hline
  \Large Our Scheme\\
  \hline
 \hspace*{-1.2cm}\textbf{Setup}. Let ($G_{1}, G_{T})$
a bilinear group.  Choose a generator $g\in G_{1}$ \\
\hspace*{-1.3cm}  and set
 $P_{pub_{1}} = g^{l}$ $\in {G_{1}}^{\star}$.
Calculate $e(g,g)=x$ and \\
 \hspace*{+1cm}  $e(g,g)^{a_{1}}=x^{a_{1}}=y_{1}$,
$e(g,g)^{a_{2}}=x^{a_{2}}=
y_{2}$,...,$e(g,g)^{a_{v}}=x^{a_{v}}=y_{v}$.
\\\hspace*{-5.8cm} (or rather $g^{a_{1}}$,
$g^{a_{2}}$,...,$g^{a_{v}}$).
\\\hspace*{+2.7cm}
$M_{pk}$= \{$G_{1},  G_{T},$ $P_{{pub}_{1}}$, $x, y_{1},
g^{a_{1}}, y_{2}, g^{a_{2}} ..., y_{v}, g^{a_{v}}$ \}, $M_{sk}$=
\{l,$a_{i}$ $/$ 1 $\leq i \leq$ v \}
\\
 \hspace*{+1.4cm} Message space is  $\{0, 1\}^{n}$, ciphertext space is ${G_{1}}^{*}$
$\times \{0, 1\}^{n} \times \{0, 1\}^{n}$.
\\\hspace*{-.9cm} \textbf{ Extract:} Given an identifer  $ID_{A}=(I_{A_{1}}, . . . , I_{A_{j}}
)$ $\in$ ${Z_{p}}^{j}$ of depth j $\leq$ v,\\ \hspace*{+.3cm} of
an entity A, public key $M_{pk}$, master key $M_{sk}$
  returns\\ \hspace*{-.3cm}
For a depth j, we have $d_{A}$=$g^{\frac{a_{1}+I_{A_{1}}+
a_{2}+I_{A_{2}}+...+ a_{j}+I_{A_{j}}}{l}}$
\\ \hspace*{+1.4cm} The private key is $(g^{\frac{a_{1}+I_{A_{1}}+
a_{2}+I_{A_{2}}+...+a_{j}+I_{A_{j}}}{l}}, g^{\frac{1}{l}}, g^{\frac{a_{j+1}}{l}},...,g^{\frac{a_{v}}{l}})$\\
\hspace*{+2.8cm}(or  $(e(g,g)^{\frac{a_{1}+I_{A_{1}}+
a_{2}+I_{A_{2}}+...+a_{j}+I_{A_{j}}}{l}}, e(g,g)^{\frac{1}{l}},
e(g,g)^{\frac{a_{j+1}}{l}},...,,e(g,g)^{\frac{a_{v}}{l}})$)\\
\\ \hspace*{+.8cm} Noting that for level j+1, choose $s_{j+1}$ $\in Z_{p}$ and calculate\\\hspace*{+.74cm}
$(g^{\frac{a_{1}+I_{A_{1}}+
a_{2}+I_{A_{2}}+...+a_{j}+I_{A_{j}}+s_{j+1}(a_{j+1})+I_{A_{j+1}}}{l}}, g^{\frac{1}{l}}, g^{\frac{a_{j+2}}{l}},...,g^{\frac{a_{v}}{l}})$\\
\textbf{ Encrypt:} Given  m $\in$ M , $ID_{A}$ and $M_{pk}$, the
following step are
formed:\\\hspace*{-4.7cm} 1. pick a random s in $Z_{q}$\\\
\hspace*{-2cm} 2.Compute
$z^{s(I_{A_{1}}+a_{1}+I_{A_{2}}+a_{2}+...+a_{j}+I_{A_{j}})}$=\\
\hspace*{+2.5cm} $= e(g,g)^{s(I_{A_{1}}+a_{1}+I_{A_{2}}+a_{2}+...+
I_{A_{j}}+a_{j})}$ =  $(x^{I_{A_{1}}+I_{A_{2}}+...+I_{A_{j}}}y_{1}y_{2}...y_{j})^{s}$ \\
\hspace*{+1.9cm}Ciphertext is $C=(g^{ls}={P_{pub_{1}}}^{s}, g^{s},
m.z^{s(I_{A_{1}}+a_{1}+I_{A_{2}}+a_{2}+...+
I_{A_{j}})})$\\
\hspace*{-1.6cm} \textbf{Decrypt:} Given  C = (u',
u'',v')$\in$$C$,
$ID_{A}$, $d_{A}$, $M_{pk}$, follow the step\\
\hspace*{-.6cm} 1. Compute e$(u',d_{A})$ and
output $m$=$\frac{v'e(u'', g^{(s_{j}-1)a_{j}})}{e(u',d_{A})}$\\
  \hline
\end{tabular}
\end{table}
\begin{flushleft}
\subsubsection*{Observation}
\begin{description}
    \item[] $\bigstar$ The private key  $(g^{\frac{a_{1}+I_{A_{1}}+
a_{2}+I_{A_{2}}+...+s_{j}(a_{j})+I_{A_{j}}}{l}}, g^{\frac{1}{l}},
g^{\frac{a_{j+1}}{l}},...,g^{\frac{a_{v}}{l}})$ = $(d_{0}, d_{1},
...d_{v-1})$ is a private key for the Entity in Hierarchy
(Children). For the user the private key will be $(
d_{0},g^{(s_{j}-1)a_{j}})$, if we are in a level j.
\end{description}
\begin{description}
    \item[] $\bigstar$ l and $a_{j}$, j $\in$ \{1,...,v \} follow a certain
level of security. What is mean that they are belonging in $2^{t}$
for a parameter t of security chosen in beginning (following for
example the requirement of NIST)
\end{description}
\subsubsection{Prove of Security} The security of our scheme is
basing on $Dl-BDHI_{WC}$ (which mean Dl-BDHI With Condition, in
the following the
condition is $g^{\alpha^{l}}=1$) assumption since:\\
\textbf{Theorem :} Suppose the (t,l,$\varepsilon$)-Decision
$BDHI_{wc}$ assumption holds in G. Then our scheme is ($t',
q_{S}$,$\varepsilon'$)-selective identity, chosen plaintext
(IND-sID-CPA) secure such that:\\ $Adv^{scheme}$($t',
q_{S}$,$\varepsilon'$) $\geq$
$Adv^{l-DBDHI_{WC}}$(t,l,$\varepsilon$)
 where t' $>$ t-O(lq $\tau$). Where $\tau$ is the time needed to make an exponentiation in the following proof:\\
\textbf{Proof.} Suppose A has advantage  in attacking our scheme.
We build an algorithm B that uses A to solve the Decision
$l-BDHI_{WC}$ problem in G. Algorithm B is given as input a random
(l+3)-tuple ($g, g^{\alpha}, g^{\alpha^{2}},..., g^{\alpha^{l-1}},
l, T) \in {G_{1}^{*}}^{l} \times Z_{q} \times G_{T}$ such that
$g^{\alpha^{l}}$=1, this input is either sampled from
$\textsl{P}_{BDHI}$ (where T = e(g, $g)^{\frac{1}{\alpha}})$
or from $\textsl{R}_{BDHI}$ (where T is uniform and independent in $G_{T}$).\\
The goal of the algorithm B is to output 1 if T = e(g,
$g)^{\frac{1}{\alpha}})$ and 0 otherwise. Algorithm B works by
interacting with A in a selective identity game as follows :\\
\textbf{Initialization.}
\begin{description}
    \item[] \hspace*{.5cm} We note for the
selective identity $ID^{*}  = ({I_{1}}^{*},...,{I_{k}}^{*}) \in
({Z_{p}})^{k}$ which algorithm A intends to attack.
 If k $<$ v,  B concatenate by 1 to have
exactly v (the depth of the hierarchy).
\end{description}
\textbf{Setup.}
\begin{description}
    \item[] \hspace*{.5cm} As algorithm A can give to B the ($g, g^{\alpha}, g^{\alpha^{2}},...,
    g^{\alpha^{l}},$ l $/$ $g^{\alpha^{l}}=1$) according to its choice. So depending  on the
    identity $ID^{*}  = ({I_{1}}^{*},...,{I_{k}}^{*})$ chosen. A
    choose an an arbitrary j from [1,k], for example j=2. He calculate ($g^{-I_{2}^{*}}$, $g^{-I_{2}^{*}\alpha},
    g^{-I_{2}^{*}\alpha^{2}},...,
    g^{-I_{2}^{*}\alpha^{l}},$ l $/$ $g^{\alpha^{l}}=1$).
 Implicitly he calculate:
    $f(\alpha)={\sum_{i=0}}^{s}\alpha^{i}$,
   $ t(\alpha)=f(\alpha)-f(0)$, also $ \frac{t(\alpha)}{\alpha}=
   \frac{f(\alpha)-f(0)}{\alpha}=f'(\alpha)$. s will be chosen according to some  requirement in phase 1. \\
   Our goal is to test if B can output the private key $d_{A} = (h^{\frac{a_{1}+a_{2}+...+a_{k}}{\alpha}+
   \frac{I_{1}-I_{1}^{*}+
   I_{2}-I_{2}^{*}+...+I_{k}-I_{k}^{*}}{\alpha}}, h^{\frac{1}{\alpha}}, h^{\frac{a_{k+1}}{\alpha}},
   h^{\frac{a_{k+2}}{\alpha}},..., h^{\frac{a_{v}}{\alpha}})$=$(d_{0},d_{1},
   d_{3},..., d_{v-2})$ for a given v
   and an identity $(I_{1}, . . . , I_{v})$\\ B first picks a random
$\gamma_{1},\gamma_{1},...,\gamma_{v} \in {Z_{p}}^{*}$ which will verify some conditions in phase 1\\
\end{description}
\textbf{Phase 1.}
\begin{description}
    \item[] \hspace*{.5cm}
 A issues up to $q_{S}$ private key queries.\\
 In the first step, choose an identity ID=$(I_{1}, . . . ,
 I_{r})$, such that $r \leq v$\\
 If $r \leq k$, he selections  only r element from $ID^{*}$
 and if $r \geq k$  the
 adversary B concatenate k (the depth of $I^{*}$) by 1 as we have seen above. \\
 To response to $d_{0}$, B can make the following step: \\
 B imagine (implicitly) that each $a_{i}$ (1 $\leq i \leq$ v)  can be writ as
 $a_{i}$=$\gamma_{i}+(-1)^{i}\alpha^{i}$ (*) \\Noting that B can make this, as he can choose a suitable
 $\gamma_{i}$ such that $g^{\alpha^{i}}=g^{a_{i}}g^{\gamma_{i}}$. We privilege to use the syntax (*), because
 $f(\alpha) g^{\alpha^{i}}$ can be not calculate-see the following\\
So $\frac{f(\alpha)-f(0)}{\alpha} \sum_{i=1}^{i=k}a_{i}$ =
$\frac{f(\alpha)-f(0)}{\alpha}\sum_{i=1}^{i=k}(\gamma_{i}+(-1)^{i}\alpha^{i})$
=
$f'(\alpha)\sum_{i=1}^{i=k}\gamma_{i}.f'(\alpha)\sum_{i=1}^{i=k}(-1)^{i}\alpha^{i}$\\
The first part $f'(\alpha)\sum_{i=1}^{i=k}\gamma_{i}$ can be
calculate easily (after exponent it by g), until the second may
not. But if we regroup it, we can find that
$f'(\alpha)\sum_{i=1}^{i=k}(-1)^{i}\alpha^{i}$ =
$\sum_{i=1}^{i=s}\alpha^{i-1}(-\alpha+\alpha^{2}-\alpha^{3}+...\alpha^{k-1}+\alpha^{k})$
=$-\sum_{i=1}^{i=s}\alpha^{i}+\sum_{i=1}^{i=s}\alpha^{i+1}-\sum_{i=1}^{i=s}\alpha^{i+2}+...
+(-1)^{k-1}\sum_{i=1}^{i=s}\alpha^{i+k-2}+(-1)^{k}\sum_{i=1}^{i=s}\alpha^{i+k-1}$.\\
To remove the overstepping $\alpha$, B must choose its s such that
s+k-1=l i.e s=l-k-1 which imply that the most long factor:
$\alpha^{i+s-1}$ is equal to 1. Thus B can calculate easily
$g^{\frac{f(\alpha)-f(0)}{\alpha}(-I_{2}^{*})
\sum_{i=1}^{i=k}a_{i}}$ = $h^{
\frac{\sum_{i=1}^{i=k}a_{i}}{\alpha}}$, (with
h=$g^{(f(\alpha)-f(0))(-I_{2}^{*})}=g^{f''(\alpha)(-I_{2}^{*})})$
which is equal to $g^{(f'(\alpha)\sum_{i=1}^{i=k}\gamma_{i})
(-\sum_{i=1}^{i=s}\alpha^{i}+\sum_{i=1}^{i=s}\alpha^{i+1}-\sum_{i=1}^{i=s}\alpha^{i+2}+...
+(-1)^{k-1}\sum_{i=1}^{i=s}\alpha^{i+k-2}+(-1)^{k}\sum_{i=1}^{i=s}\alpha^{i+k-1})(-I_{2}^{*})}$
= $g^{(f'(\alpha)\sum_{i=1}^{i=k}\gamma_{i})(- \alpha -
\alpha^{3}-...(-1)^{k}\alpha^{s+k-1})(-I_{2}^{*})}$.\\
For the second part: R = $h^{
   \frac{I_{1}-I_{1}^{*}+
   I_{2}-I_{2}^{*}+...+I_{k}-I_{k}^{*}}{\alpha}}$. To output the
   exact key of ID at which all elements of ID operate in $d_{0}$,
   all the $I_{i}$ chosen will be different from $I_{2}^{*}$. And
to  benefit
   from $f''(\alpha)$,  all  $I_{i}$ (for all 1 $\leq i \leq$ r) of the requested identity ID,  will be such
   that: $I_{i} \neq n I_{2}^{*}$ from each to
   other and this for  n $\in$ \textsl{N}.  Because
   he wouldn't obtain $f''(\alpha)$, but he may obtain another
   $f'''(\alpha)$. \\
\textbf{Observation}\\
   A can
   choose ($g^{-I_{k}^{*}}$, $g^{-I_{2}^{*}\alpha^{2}},
    g^{-I_{4}^{*}\alpha^{4}},...,
    g^{-I_{k-1}^{*}\alpha^{k-1}},$ l $/$ $g^{\alpha^{l}}=1$)instead of ($g^{-I_{2}^{*}}$, $g^{-I_{2}^{*}\alpha},
    g^{-I_{2}^{*}\alpha^{2}},...,
    g^{-I_{2}^{*}\alpha^{l}},$ l $/$ $g^{\alpha^{l}}=1$) ( we treat this later i.e only with $I_{2}^{*}$ to simplify  the
    proof). So if B make a research exhaustive to know the exact place of $I_{i}^{*}$ for $1 \leq i \leq v $,
    he need at most doing v research, which cost (v!), as v can be
    great. So for all $1 \leq i \leq v $ the $I_{i} \neq n I_{2}^{*}$ $\forall$ n $\in$ \textsl{N}. And this is an ideal case.\\
\vspace*{.4cm} To calculate R, B will calculate firstly $d_{1}$.
And to do
    it, B can calculate
    $\frac{f(\alpha)-f(0)}{\alpha}=f'(\alpha)$. After he calculate
    $g^{\frac{f(\alpha)-f(0)}{-\alpha}(-I_{2}^{*}})=g^{\frac{f''(\alpha)(-I_{2}^{*})}{\alpha}}=g^{f'(\alpha)(-I_{2}^{*})}=h^{\frac{1}{\alpha}}=d_{1}$.
    With this, B can calculate easily R, as he exponents only
    with $I_{1}-I_{1}^{*}+
   I_{2}-I_{2}^{*}+...+I_{k}-I_{k}^{*}$.\\
Now to calculate $d_{3},d_{4}...,d_{v-2}$, we have respectively
the coefficients $\alpha, \alpha^{2}, \alpha^{3},...,
\alpha^{s+v-1}$ after a product of $a_{k+1}$...,$a_{v}$ with
$f'(\alpha)$. Effectively, all j overstepping l i.e l=j-x their
$\alpha^{j}=\alpha^{x}$, with x$<$l    \hspace*{.2cm}   $\Box$\\
Thus with this manner B can responds to the private key $d_{A} =
(h^{\frac{a_{1}+a_{2}+...+a_{k}}{\alpha}+
   \frac{I_{1}-I_{1}^{*}+
   I_{2}-I_{2}^{*}+...+I_{k}-I_{k}^{*}}{\alpha}}, h^{\frac{1}{\alpha}}, h^{\frac{a_{k+1}}{\alpha}},
   h^{\frac{a_{k+2}}{\alpha}},..., h^{\frac{a_{v}}{\alpha}})$
\end{description}
\textbf{Challenge.}
\begin{description}
    \item[] \hspace*{.4cm} A outputs two messages $M_{0}, M_{1} \in G_{1}$. Algorithm B
picks a random bit b $\in$ \{0,1\} and a random l' $\in$
${Z_{p}}^{*}$. It responds with the ciphertext prepared as
follow:\\
He have  ${g^{(f(\alpha)-f(0))(-{I_{2}}^{*})}}^{s}=h^{\frac{s}{\alpha}.\alpha}=h^{l'\alpha}$ = $c_{1}$, with l'=$\frac{s}{\alpha}$\\
And
$c_{2}$=$MT_{h}^{s(a_{1}+a_{2}+...+a_{k}+I_{1}^{*}+I_{2}^{*}+...+I_{k}^{*})}=
T_{h}^{s(a_{1}+a_{2}+...+a_{k}+I_{1}^{*}+I_{2}^{*}+...+I_{k}^{*})}$ \\
So if $T_{h}=e(h,h)^{\frac{1}{\alpha}}$ he will have
$e(h,h)^{\frac{s}{\alpha}(a_{1}+a_{2}+...+a_{k}+I_{1}^{*}+I_{2}^{*}+...+I_{k}^{*})}=c_{2}=e(h,h)^{l'(a_{1}+
a_{2}+...+a_{k}+I_{1}^{*}+I_{2}^{*}+...+I_{k}^{*})}$\\
And he combine
CT=$(c_{1},c_{2})=(h^{l'\alpha},e(h,h)^{l'(a_{1}+a_{2}+...+a_{k}+I_{1}^{*}+I_{2}^{*}+...+I_{k}^{*})})$
which is a
valid ciphertext under $ID^{*}$\\
If $T_{h}$ is uniform in $G_{1}$, then CT is independent of the
bit b.\\
\end{description}
 \textbf{Phase 2.} \begin{description}
    \item[] \hspace*{.4cm} A issues more private key queries, for a total
of at most $q_{S}$ $<$ q. Algorithm B responds as before.
\end{description}
\textbf{Guess.} \begin{description}
    \item[] \hspace*{.4cm} Finally, A outputs a guess b' $\in$ \{0, 1\}. If b =
    b'
then B outputs 1 meaning T = $e(g, g)^{\frac{1}{\alpha}}$.
Otherwise, it outputs 0 meaning T $\neq$ $e(g,
g)^{\frac{1}{\alpha}}$.
\end{description}
When the input $l+2$-tuple is sampled from $\textsl{P}_{BDHIP}$
(where T = $e(g,g)^{\frac{1}{\alpha}}$) then A's view is identical
to its view in a real attack game and therefore A must satisfy
$|$Pr[b = $b'$] - 1/2$|$ $>$ $\varepsilon$. On the other hand,
when the input $l+2$-tuple is sampled from $\textsl{R}_{BDHIP}$
(where T is uniform in $G_{T}$) then Pr[b = $b'$] = 1/2.
Therefore, with g uniform in
$G_{1}$, T uniform in $G_{T}$ we have that:\\
$ \left |%
     \begin{array}{ll}
    \hbox{Pr [ g, $g^{\alpha}, g^{\alpha^{2}},..., g^{\alpha^{l-1}}$, l, \^e($g,g)^{\frac{1}{\alpha}}$ ]-
    Pr [g, $g^{\alpha}, g^{\alpha^{2}},..., g^{\alpha^{l}}$, l, T ] } \\
\end{array}%
\right |$ $\geq$
$ \left |%
     \begin{array}{ll}
    \hbox{($\frac{1}{2}\pm \varepsilon)-\frac{1}{2}$=$\varepsilon$}
\end{array}%
\right |.$ \hspace*{2cm} $\Box$ \hspace*{-1cm}
\subsubsection{Discussion} Our first discussion will be about the
problem used in the proof, which is $Dl-BDHI_{WC}$. We have
considers in the above that: l $>>$ v (v is the depth of the
hierarchy). But, this can make our proposition vulnerable to the
cryptanalysis of Cheon [20] by comparison with $Dl-wBDHI^{*}$ in
[17]. As in this latter, l $\leq$ v (v the depth of the
hierarchy), since in the [20] cheon prove that the strong
Diffie-Hellman problem has a complexity reduction $O(\sqrt{l})$ by
comparison with PDL. So while k is great, while it will be easy to
be cryptanalysis. To avoid this, we propose to consider l=v+l', we
can use so $\alpha^{l'}=\beta$ instead of $\alpha$ to reduce the
problem from $Dl-BDHI_{WC}$ to $Dv-BDHI_{WC}$ and even we can make
less of this.\\We note that the relationship between the problem
used is: $ l-BHIP \longrightarrow^{1} l-wBDHI^{*}
\longrightarrow^{2} l-BDHI_{WC}$ (so: $ Dl-BHIP
\longrightarrow^{1} Dl-wBDHI^{*} \longrightarrow^{2}
Dl-BDHI_{WC}$). The relation 1 was proven in [17], until 2 is easy
to be proven.\\Even if [17], is basing on a strong problem of
Diffie Hellman compared to our (this may be linked to the use of
asymmetric pairing). But [17] has two weakness, which are the
obliged use of the selection identity in the study of simulation
in $Z_{p}^{*}$ instead of $Z_{p}$ as with our. This limit the
selection of the identity to be challenged, since we couldn't use
any were the bit 0. More than that the [17] does not support
$s^{+}ID-CPA$, by contrast our scheme is like BB1 support this
notion. According to [14] to render [17] $s^{+}ID-CPA$, the
authors make a simple modification. Its proof yields a
multiplicative security degradation by a factor of v, where v is
the maximum number of levels in the HIBE. And to not obtain this
degradation the authors add v-k factors or rather
$(v-k)Exp_{G_{1}}$ in the original scheme (v is the maximum  depth
of the Hierarchies, until k is the depth
of the identity selected $ID^{*}$)\\
By contrast with our scheme we don't need this, because our scheme
is $s^{+}ID-CPA$ and it offer a competitive to [17]\\To see this
we count in the following the complexity of BB1, BBG, and our
scheme:\\
\begin{tabular}{|c|c|c|c|}
  \hline
  & $Extract_{user}$ level k  & Encrypt & Decrypt  \\
  \hline
  BB1 &$(2k+3)Exp_{G_{1}}$   & $(2k+1)Exp_{G_{1}}+1Exp_{G_{T}}$ & $(k+1)pairing+kMul_{G_{T}}$  \\
  \hline
   BBG & $3 Exp_{G_{1}}$ & $(k+2)Exp_{G_{1}}+1Exp_{G_{T}}$  &  2pairing  \\
   \hline
   Our & $2 Exp_{G_{1}}$ or $2 Exp_{G_{T}}$ & $(k+2)Exp_{G_{T}}+2Exp_{G_{1}}$ &  2 pairing+$1Mul_{G_{T}}$+$1Exp_{G_{1}}$  \\
\hline
\end{tabular}\\
In this table we wouldn't take into account some complexity (like
division of pairing, multiplicity by $y_{1}y_{2}...y_{k}$ in our
scheme, multiplicity by $g_{3}$ in BBG...)
\\ According to this
table our scheme  is more efficient by comparison with BB1 and
with even BBG. Because, $Exp_{G_{T}}$ which we count it as
$Exp_{Z_{p^{k'}}}$ (in the finite field) is small than
$Exp_{G_{1}}$ (i.e in
curve elliptic).\\
This efficient is visible in Extract, and Encrypt (for the two
scheme BB1 and BBG). For the Decrypt we have a little overstepping
by comparison with BBG, but because of what we seen in the highest
(in the point of view security), our scheme is so more efficient.
\subsection{Application}
\subsubsection{Overview on Forward Encryption} In [13] Canetti et al propose a
forward-secure encryption scheme in the standard model basing on
[16]. The (fs-HIBE) scheme allows each user in the hierarchy to
refresh his or her private keys periodically while keeping the
public key the same. Using this, so even if there are any were a
compromise
 of long-term keys it does not permit the compromise of the past session
keys and therefore past communications. Since exposure of a secret
key corresponding to a given interval does not enable an adversary
to break the system  for any prior time
period. For more detail, we send the interested to [13][33].\\
To admit a succeed  Forward Security, the following requirements
will be realizing:\\
\begin{description}
    \item[] \hspace*{.5cm}- New users would be able to join the hierarchy and receive secret
keys from their parent nodes at any time.
    \item[]\hspace*{.5cm} - The encryption  does not
require knowledge of when a user or any of his ancestors joined
the hierarchy, we call this joining-time-oblivious. So the sender
can encrypt the message as long as he knows the current time and
the ID-tuple of the receiver, along with the public parameters of
the system.
    \item[]\hspace*{.5cm}- The scheme should be forward-secure.
    \item[]\hspace*{.5cm} - Refreshing secret keys can be carried out autonomously, that is, users can refresh
their secret keys on their own to avoid any communication overhead
with any PKG.
\end{description}
Eventually jointing [13] and [16] can give a scheme which can not
verify these requirements. For more detail see [33]. To over come
this the authors in [33], have proposed a scheme (basing in [13])
which conserve all these requirements, but they use only HIBE of
[16], which give a heavy scheme.\\ In the following we give a
version at which we use our syntax of an HIBE (we declared it
only). This  reduce the complexity, but because of some
circumstance, we wouldn't give in this article it's proof of
security. We let it, in the future work and to the interested.
\subsubsection*{Implementation: Declaration}
Firstly we note $sk_{w,(ID_{1},...,ID_{v})}$: a node key
associated with some prefix w of he bit representation of a time
period i and a tuple $(ID_{1}, . . . , ID_{v})$.\\
$SK_{i,(ID_{1},...,ID_{v})}$: Key associated with time i and an
ID-tuple $(ID_{1}, . . . , ID_{v})$. It consists of sk keys as
follows: $SK_{i,(ID_{1},...,ID_{v})}$ =
$\{sk_{i,(ID_{1},...,ID_{v})}$, $sk_{w_{1},(ID_{1},...,ID_{v})}$:
$w0$ is a prefix of i\}. With W0 and W1 represent respectively node right and node left.\\
\textbf{Setup}$(1^{k },N = 2^{l})$ \\
    \item[] The root PKG with $ID_{1}$ does the
following:\\
\begin{enumerate}
    \item \textit{IG} is run to generate groups $G_{1},G_{T}$ of order q
and bilinear map \^e.
    \item A random generator g of $G_{1}$ is selected

 \item  $P_{pub_{1}} = g^{l}$ $\in {G_{1}}^{\star}$.
\item Calculate $e(g,g)=x$,  $e(g,g)^{a_{1}}=x^{a_{1}}=y_{1}$,
$e(g,g)^{a_{2}}=x^{a_{2}}=
y_{2}$,...,$e(g,g)^{a_{v}}=x^{a_{v}}=y_{v}$.
\\ (or rather $g^{a_{1}}$,
$g^{a_{2}}$,...,$g^{a_{v}}$).\\ $M_{pk}$= \{$G_{1},  G_{T},$
$P_{{pub}_{1}}$, $x, y_{1}, g^{a_{1}}, y_{2}, g^{a_{2}} ...,
y_{v}, g^{a_{v}}$ \}, $M_{sk}$= \{l,$a_{i}$ $/$ 1 $\leq i \leq$ v
\}
\end{enumerate}
The following algorithm is a helper method, it is called by the
Setup and Upd algorithms.\\
\textbf{CompNext}$(sk_{w,h},w, (ID_{1} . . . ID_{v}))$
\begin{description}
    \item[] \hspace*{.5cm} It takes a
secret key $sk_{w,v}$, a node w, and an ID-tuple, and outputs keys
$sk_{(w0),v}, sk_{(w1),v}$ for time nodes w0 and w1 of $(ID_{1}
... ID_{v})$.
\begin{enumerate}
    \item Parse w as $w_{1} . . .w_{d}$, where $|w|$ = d. Parse ID-tuple as $ID_{1}, . . . , ID_{v}$. Parse
$sk_{w,h}$ associated with time node w, for all 1$\leq$ k $\leq$ d
and 1 $\leq$ j $\leq$ v.
    \item Choose random $s_{(d+1),j} \in Z_{q}$ for all $1 \leq j \leq h$.
    \item Set $S_{(w0),v}$ = $(g^{\frac{a_{d+1,1}+w0 \circ I_{A_{1}}+
a_{d+1,2}+w0 \circ I_{A_{2}}+...+a_{d+1,j-1}+w0 \circ
I_{A_{j-1}}+s_{ d+1,j}(a_{d+1,j})+w0 \circ I_{A_{j}}}{l}},
g^{\frac{1}{l}}, g^{\frac{a_{d+1,j+1}}{l}},...,g^{\frac{a_{d+1,v}}{l}})$\\
$S_{(w1),h}$ = $(g^{\frac{a_{d+1,1}+w1 \circ I_{A_{1}}+
a_{d+1,2}+w1 \circ I_{A_{2}}+...+a_{d+1,j-1}+w1 \circ
I_{A_{j-1}}+s_{d+1,j}(a_{d+1,j})+w1 \circ I_{A_{j}}}{l}},
g^{\frac{1}{l}}, g^{\frac{a_{d+1,j+1}}{l}},...,g^{\frac{a_{d+1,v}}{l}})$\\
    \item Erase $s_{(d+1),j}$ for all $1 \leq j \leq v$.
\end{enumerate}
\end{description}
\textbf{KeyDer($SK_{i,(v-1)}, i, (ID_{1} . . . ID_{v}))$}
\begin{description}
    \item[]\hspace*{.7cm} Let $E_{h}$ be an entity that joins the hierarchy during the time
period i $<$ N - 1 with ID-tuple $(ID_{1}, . . . , ID_{v}).
E_{h}'s$ parent generates $E_{v}'s$ key $SK_{i,v}$ using its key
$SK_{i,(v-1)}$ as follows:
\begin{enumerate}
    \item Parse i as $i_{1} ... i_{l}$ where l = $log_{2} N$.
Parse $SK_{i,(n-1)}$ as $(sk_{i,(v-1)},$  \{
${sk_{(i|_{k-1}1),(v-1)}}]_{i^{k}}$ \}=0).
    \item For each value $sk_{w,(v-1)}$ in $SK_{i,(v-1)},$ $E_{v}'s$ parent does the following to
generate $E_{h}'s$ key $sk_{w,v}$: (a) Parse w as $w_{1} . . .
w_{d}$, where $d \leq l$, and parse the secret key $sk_{w,(v-1)}$
as $(S_{w,(v-1)},,
g^{\frac{1}{l}}, g^{\frac{a_{w,v}}{l}})).$\\
(b) Choose random $s_{k,v}\in Z_{q}$ for all 1 $\leq k \leq$ d.
Recall that $s_{k,j}$ is a shorthand for $s_{w|_{
k}}$,$(ID_{1}...ID_{j}$) associated with time node $w|_{k}$ and
tuple $(ID_{1} . . . ID_{j})$.\\
(c) Set the child entity $E_{v}$'s secret point $S_{w,v}$
=$g^{\frac{a_{1,1}+w|_{k} \circ I_{A_{1}}+ a_{2,2}+w|_{k} \circ
I_{A_{2}}+...+a_{j-1,j-1}+w|_{k} \circ I_{A_{j-1}}+s_{
d+1,j}(a_{j,j})+w|_{k} \circ I_{A_{j}}}{l}}$.\\
 \item [] $E_{h}'s$ parent sets $SK_{i,h} = (sk_{i,h}$, \{$sk_{(i|_{k-1}1),h} \}_{{i_{k}=0})}$, and erases all other information.
\end{enumerate}
\end{description}
\textbf{Upd}($SK_{i,h}, i + 1, (ID_{1} ... ID_{v})$) (where i $<$
N -1)
\begin{description}
    \item[] At
the end of time i, an entity (PKG or individual) with ID-tuple
($ID_{1}, . . . , ID_{v}$) does the following to compute its
private key for time i + 1, as in the fs-PKE scheme
[].\\
\begin{enumerate}
    \item Parse i as $i_{1} ... i_{l}$, where $|i|$ = l. Parse $SK_{i,v}$ as $(sk_{(i|_{l}),v},
    \{sk_{(i|_{k-1}1),v}\}_{i_{k}}=0)$. Erase $sk_{i|_{l},h}$.
    \item We distinguish two cases. If $i_{l} = 0$, simply output the remaining keys as the
key $SK_{(i+1),v }$ for the next period for ID-tuple ($ID_{1}, . .
. , ID_{h}$). Otherwise, let $\widetilde{k}$ be the largest value
such that $i_{\widetilde{k}}$ = 0 (such $\widetilde{k}$ must exist
since i $<$ N - 1). Let i' = $i|_{\widetilde{k}-1}1$. Using
$sk_{i',h}$ (which is included as part of $SK_{i,v}$), recursively
apply algorithmCompNext to generate keys $sk_{(i'0^{d}1),v}$ for
all\\ 0 $\leq d \leq l-\widetilde{k}-1$, and
$sk_{(i'0^{d-\widetilde{k}},v)}$. The key
$sk_{(i'0^{d-\widetilde{k}},v)}$ will be used for decryption in
the next time period i+1, the rest of sk keys are for computing
future keys. Erase $sk_{i',v}$ and output the remaining keys as
$SK_{(i+1),v}$.
\end{enumerate}
\end{description}
\textbf{Enc(i, $(ID_{1},..., ID_{v}),M)$ (where M $\in
\{0,1\}^{n}$)}
\begin{description}
\item[]
\begin{enumerate}
\item Parse i as $i_{1} ... i_{l}$ \item Denote $P_{k,j}$ =
$H_{1}(i|_{k} \circ ID_{1} ... ID_{j})$ for all $1 \leq k \leq l$
and $1 \leq j \leq h$. \item pick a random s in $Z_{q}$ \item
Compute $z^{s(a_{|_{2},1}+i|_{2} \circ ID_{1}+...+a_{|_{j},1}+
i|_{j} \circ ID_{1}+ a_{|_{1},1}+i|_{1} \circ
ID_{1}+...+a_{|_{1},j}+ i|_{1} \circ ID_{1} ...
ID_{j}+...+a_{|_{j},1}+i|_{j} \circ ID_{1}+...+a_{|_{j},j}+ i|_{j}
\circ ID_{1}...ID_{j})}$=

$e(g,g)^{s(a_{|_{2},1}+i|_{2} \circ ID_{1}+...+a_{|_{j},1}+ i|_{j}
\circ ID_{1}+ a_{|_{1},1}+i|_{1} \circ ID_{1}+...+a_{|_{1},j}+
i|_{1} \circ ID_{1} ... ID_{j}+...+a_{|_{j},1}+i|_{j} \circ
ID_{1}+...+a_{|_{j},j}+ i|_{j}
\circ ID_{1}...ID_{j})}$\\
 Ciphertext is C=($g^{ls}={P_{pub_{1}}}^{s},
g^{s}$, $m.z^{s(a_{|_{2},1}+i|_{2} \circ ID_{1}+...+a_{|_{j},1}+
i|_{j} \circ ID_{1}+ a_{|_{1},1}+i|_{1} \circ
ID_{1}+...+a_{|_{1},j}+ i|_{1} \circ ID_{1} ...
ID_{j}+...+a_{|_{j},1}+i|_{j} \circ ID_{1}+...+a_{|_{j},j}+ i|_{j}
\circ ID_{1}...ID_{j})})$
\end{enumerate}
\end{description}
 \textbf{Decrypt:} Given  C =
(u',u'',v')$\in$$C$, $ID_{A}$, $d_{A}$, $M_{pk}$, follow the step
\begin{description}
\item[]
\begin{enumerate}
 \item Parse i as $i_{1} ...i_{l}$. Parse $SK_{i,h}$ associated with the ID-tuple as $(sk_{i,h},
 \{sk_{(i|_{k-1}1),h}\}i_{k}$=0).
\item Compute e$(u',d_{A})$ and
output $m$=$\frac{v'e(g^{s}, g^{s_{j}a_{|_{j},1}+...+a_{|_{j},j}})}{e(u',d_{A})}$\\
\end{enumerate}
\end{description}
\subsection*{Comparison} To see the efficiency of our scheme (and
BBG) in forward scheme we make the following comparison.\\
\begin{tabular}{|c|c|c|}
  \hline
   & fs-HIBE [33] & fs-with our \\
  \hline
  Key derivation time & O(v log N) & O((v-k) logN) \\
 Encryption time & O(v log N) & O(v log N) \\
 Decryption time & O(v log N) & O(k+log N) \\
 Key update time & O(v) & O(v-k) \\
  Ciphertext length & O(v log N) & O(3 log N) \\
 Public key size & O(v + log N) & O(v + log N) \\
  Secret key size & O(v log N) & O((v-k) log N) \\
\hline
\end{tabular}\\
k is the hierarchy children considered.\\
N is the total number of the time periods.\\
v is depth of the hierarchy. \\

\subsection{Construction of CCA2}
This section is reserved to signal the technique to be used to
obtain a CCA2 from CPA.\\
To render CPA a CCA2, there are some techniques:\\
For an IBE or HIBE with random oracle we can use the two method
given by Fujusiki Okamoto [34]\\
For an IBE or HIBE without random oracle, there are also
two techniques:\\
That's of [13] at which we use one-time signature.\\
That's of [35] at which we add a MAC.\\
So using one of these last technique can render our scheme CCA2
secure.
\section{Conclusion}
In these papers, we have study the competition between the
best-known cryptosystems of the  cryptography  IBE. Our approach
is more accurate than the only method  made in this direction of
Boyen. Even if we follow a very simple strategy but it is so
effective to clarify the cryptosystems that deserve a standardized
participation. We concluded that the pattern of Boneh and Franklin
in the field of RO, is the most effective, but we recommend using
one of Skai Kasarah since Boneh and Franklin projects into an
elliptic curve which limit the selection of curve, it may so pose
a problems of security. And we note that unlike the results of
Boyen the BB1 is late compared to others. In general we can say
that the scheme of Water is the most preferable as it is traced in
the domain of SM, more it has an important classification.
Following the criteria considered SK and BF are the most
helpful.\\ This study is very useful to cryptographers, because we
surveying the very recents recherches in IBE. More we shows the
weakness and strength of every cryptosystem in competition, which
can facilitate to make an improvement to admit a more practical
cryptosystems.\\
More than that, we have presented two efficient schemes in the
model selective ID and without random oracle (which is our second
contribution behind this work). With a little change in the
schemes of Boneh and Boyen we get a more efficient schemes. The
change is make in BB2 (change $\frac{1}{s+ID}$ by $\frac{1}{s}$),
which permit to eliminate the use of two pairing in the Decrypt of
IBE and, more the resulting scheme is traced in the approach of
commutative Blinding. Effectively as it is presented in this
article, the complexity of our scheme is less than that of BB1
(version IBE) and even than that of BB2. More than that, we have
based our prove of security in $Dk^{-}$-BDHIP which is an
efficient problem than Dk-BDHIP used by BB2, since with this
latter, k is linked essentially to the numbers of identity to be
challenged. By contrast, with our we are not, any $k^{-}$ can
serve us, we can take as title of example $k^{-}$=2, which make
$Dk^{-}$-BDHIP in competition with DBDHP (D1-BDHIP) used by BB1.
In other part, using our syntax of IBE in HIBE and using the
technique of BBG (Boneh Boyen Goh) we get a more efficient HIBE
than BB1 and BBG. The efficiency by comparison with BB1, is
clearly seen in complexity. With our proposition, the technique of
BBG will be more efficient. Because, with our proposition the
complexity will be reduced. More than that, our HIBE support
$s^{+}$-ID (which require a degradation by v in the studies of
simulations) and we can not demand that the identity to be
challenged will be in $Z_{q}^{*}$ as with BBG. This render BBG
more restricted, as we are are not free to choose the identity to
be challenged. Using our proposition in some applications like
Forward Encryption make them
more efficient.\\
Thus, during all these papers, we have presented an efficient IBE
and HIBE without random oracle. With a little change in BB2 we
obtain an efficient schemes than BB1 and BB2, which are considered
until 2011 (Journal of Cryptology) as the most efficient schemes
in the model selective ID and without random oracle.
\end{flushleft}
\subsection*{Acknowledge}
We would like to thank the head of our laboratory Mr.Aboutajdinne
Driss.


\begin{thebibliography}{1}
\bibitem{} A. Shamir. Identity-based cryptosystems
and signature schemes. In G. R. Blakley and David Chaum, editors,
Advances in Cryptology - CRYPTO'84, volume 196 of Lecture Notes in
Computer Science, pages 47-53. Springer-Verlag, 1985.
\bibitem{}D. Boneh and M. Franklin. Identity based encryption from
the Weil pairing. SIAM Journal on Computing, 32(3):586-615, 2003.
\bibitem{}D. Boneh and X. Boyen. Efficient selective-ID secure identity
based encryption without random oracles. In Christian Cachin and
Jan Camenisch, editors, Advances in Cryptology - EUROCRYPT 2004,
volume 3027, pages 223-238, 2004.
\bibitem{IBEWP}R. Sakai and M. Kasahara. ID based cryptosystems with pairing on
elliptic curve. Cryptology ePrint Archive, Report 2003$/$054.
\bibitem{} B. Waters. Efficient
identity-based encryption without random oracles. In Ronald
Cramer, editor, Advances in Cryptology - EUROCRYPT 2005, volume
3494 of Lecture Notes in Computer Science, pages 114-127.
Springer-Verlag, 2005.
\bibitem{}Gentry. Practical identity-based encryption without random oracles. In
Serge Vaudenay, editor, Advances in Cryptology - EUROCRYPT 2006,
volume 4004 of Lecture Notes in Computer Science, pages 445-464.
Springer-Verlag, 2006.
\bibitem{} E. Kiltz, Y. Vahlis. CCA2 Secure IBE: Standard Model Efficiency
through Authenticated Symmetric Encryption.  CT-RSA 08, Lecture
Notes in Computer Science Vol. , T. Malkin ed., Springer-Verlag,
2008.
\bibitem{}E. Kiltz. Chosen-ciphertext secure identity-based encryption in the standard model with short
ciphertexts. Cryptology ePrint Archive, Report 2006/122, 2006.
\bibitem{} IEEE P1363.3 Committee. IEEE 1363.3 - standard for identity-based cryptographic techniques
using pairings. http://grouper.ieee.org/groups/1363/, April 2007.
\bibitem{} X. Boyen. The BB1 identity-based cryptosystem: A standard for
encryption and key encapsu- lation. Submitted to IEEE 1363.3, aug
2006. http://grouper.ieee.org/groups/1363/.
\bibitem{} X. Boyen. A tapestry of identity-based encryption: Practical frameworks compared. International
Journal of Applied Cryptography, 1(1):3-21, 2008.
\bibitem{} M. Bellare, A. Desai, D. Pointcheval, and Ph
Rogaway. Relations among notions of security for public-key
encryption schemes, volume 1462 Lecture Notes in Computer Science,
pages 26-45. Springer-Verlag, 1998
\bibitem{IBEWP} R. Canetti, S. Halevi, and J. Katz. Chosen-ciphertext security from identity-based encryption.
In Advances in Cryptology—EUROCRYPT, volume 3027 of LNCS, pages
207–22. Springer-Verlag.
\bibitem{} Sanjit Chatterjee and Palash Sarkar. Constant Size Ciphertext HIBE in the Augmented Selective-ID Model
and its Extensions. IACR eprint archive report 084/2007.
\bibitem{IBEWP} J. Horwitz and B. Lynn. Toward hierarchical identity-based
encryption. In Lars R. Knudsen, editor, Advances in Cryptology -
EUROCRYPT 2002, volume 2332 of Lecture Notes in Computer Science,
pages 466-481. Springer-Verlag, 2002.
\bibitem{IBEWP}C. Gentry and A. Silverberg. Hierarchical ID-based
cryptography. In Yuliang Zheng, editor, Advances in Cryptology -
ASIACRYPT 2002, volume 2501 of Lecture Notes in Computer Science,
pages 548-566. Springer-Verlag, 2002.
\bibitem{IBEWP} D. Boneh, X. Boyen, and Eu-Jin Goh. Hierarchical identity
based encryption with constant size ciphertext. In Ronald Cramer,
editor, Advances in Cryptology - EUROCRYPT 2005, volume 3494 of
Lecture Notes in Computer Science, pages 440-456. Springer-Verlag,
2005.
\bibitem{IBEWP}D. Boneh and X. Boyen. Efficient selective-ID secure identity
based encryption without random oracles. Journal of Cryptology
(JOC), 24 (4):659-693, 2011. Extended abstract in proceedings of
Eurocrypt 2004, LNCS 3027, pp. 223-238, 2004 i.e [5]
\bibitem{IBEWP}L. Chen,
Zh. Cheng $\shortparallel$ Security Proof of Sakai-Kasahara's
Identity-Based Encryption Scheme $\shortparallel$  In Proceedings
of Cryptography and Coding 2005.
\bibitem{IBEWP}J. Cheon. Security analysis of the strong Diffie-Hellman
problem. In Serge Vaudenay, ed- itor, EUROCRYPT 2006, volume 4004
of LNCS, pages 1-11. Springer-Verlag, Berlin, Germany, May / June
2006.
\bibitem{} M. Bellare and P. Rogaway. Random oracles are practical: a
paradigm for designing e±cient protocols. In Proceedings of the
First Annual Conference on Computer and Communications Security,
ACM, 1993.
\bibitem{} Ga\"{e}tan Leurent and Phong Q.
Nguyen. How risky is the random-oracle model? In Halevi [18],
pages 445{464}.
\bibitem{IBEWP} D. Galindo. A separation between selective and full-identity security notions
for identity-based encryption Available on: IACR eprint archive.
\bibitem{OIPBC}L Martin. "Introduction To Identity Based
Encryption". Available at:
http://www.artechhouse.com/GetBlob.aspx?strName=Martin-238-CH04.pdf
\bibitem{IBEWP} D. Galindo $\shortparallel$ Boneh-Franklin identity based encryption revisited $\shortparallel$. In Proceedings of
the 32nd International Colloquium on Automata, ICALP 2005.
\bibitem{IBEWP} S. Galbraith, K. Paterson, and N. Smart.
Pairings for cryptographers. Discrete Applied Mathematics,
156(16):3113-3121, 2008.
\bibitem{IBEWP} S. Marie-Aude $\shortparallel$ Etude de la Primalit\'{e} motiv\'{e}e par le besoin
de Nombres Premiers dans le Chiffrement RSA $\shortparallel$ sur
le site: http://www-magistere.u-strasbg.fr/IMG/pdf/MASteineur.pdf
\bibitem{IBEWP}H.Cohen,
G. Frey. Handbook of Elliptic and Hyperelliptic Curve
Cryptography.

\bibitem{IBEWP} Tetsuya Izu and Tsuyoshi Takagi. Efficient Computations of the Tate
Pairing for the Large MOV Degrees. In ICISC 2002, volume 2587 of
Lecture Notes in Computer Science, pages 283-297. Springer Verlag,
2003.

\bibitem{IBEWP} Nadia El Mrabet, Arithm\'{e}tique des couplages, performance et r\'{e}sistance
aux attaques par canaux cach\'{e}s. December 2009, Th\`{e}se.

\bibitem{IBEWP}N. Koblitz and A. Menezes. Pairing-based cryptography at high
security levels. In Nigel P. Smart, editor, Cryptography and
Coding, volume 3796 of Lectures Notes in Computer Science, pages
13-36, Berlin, Heidelberg, 2005. Springer-Verlag.

\bibitem{} Galindo and Ichiro Hasuo. Security Notions for Identity Based
Encryption. available on: http://eprint.iacr.org/2005/253

\bibitem{IBEWP} D.(Daphne) YAO, N.FAZIO , Y.DODIS  and A.LYSYANSKAYA.
Forward-Secure Hierarchical IBE with Applications to Broadcast
Encryption. Chapiter of book: Identity-Based Cryptography, in M.
Joye and G. Neven (Editors). 2009.
\bibitem{IBEWP} E. Fujisaki and T. Okamoto. Secure integration of asymmetric and
symmetric encryption schemes. In Proceedings of Advances in
Cryptology - CRYPTO '99, LNCS 1666, pp. 535-554, Springer-Verlag,
1999.
\bibitem{IBEWP} D. Boneh and J. Katz. Improved efficiency for CCA-secure
cryptosystems built using identity based encryption. Submitted for
publication, 2004.


\end{thebibliography}
\end{document}